# A Comprehensive Review of Quantum Circuit Optimization: Current Trends and Future Directions

**Krishnageetha Karuppasamy *, Varun Puram, Stevens Johnson and Johnson P. Thomas**

Department of Computer Science, Oklahoma State University, Stillwater, OK 74078, USA;
vpuram@okstate.edu (V.P.); stevens.johnson@okstate.edu (S.J.); johnson.thomas@okstate.edu (J.P.T.)
* Correspondence: kkarupp@okstate.edu

**Abstract:** Optimizing quantum circuits is critical for enhancing computational speed and mitigating errors caused by quantum noise. Effective optimization must be achieved without compromising the correctness of the computations. This survey explores recent advancements in quantum circuit optimization, encompassing both hardware-independent and hardware-dependent techniques. It reviews state-of-the-art approaches, including analytical algorithms, heuristic strategies, machine learning-based methods, and hybrid quantum-classical frameworks. The paper highlights the strengths and limitations of each method, along with the challenges they pose. Furthermore, it identifies potential research opportunities in this evolving field, offering insights into the future directions of quantum circuit optimization.

**Keywords:** quantum computing; quantum circuit optimization; quantum machine learning; hybrid algorithm; quantum circuit layouts

## 1. Introduction

Quantum computing has the potential to change the landscape of computing. Tech behemoths like Google, Microsoft, and IBM have invested millions of dollars in quantum research. Quantum information science is now at the same stage of development as a classical computer was in the 1950s. Quantum computing adopts quantum mechanics concepts such as superposition, entanglement, and interference to improve the performance of solving hard problems that are difficult to solve by even the most powerful classical supercomputers.

The quantum factorization algorithm [1] developed by Peter Shor takes polynomial time compared to exponential time to execute in a classical computer. Searching an element in unstructured databases takes at least O($n$) in a classical algorithm, but the Grovers quantum search algorithm [2] takes only O($\sqrt{N}$). In recent years, numerous applications have emerged leveraging the quantum framework, including quantum optimization for 6G wireless communications [3] and significant advancements in machine learning [4–7].

The emergence of advanced quantum devices, including Google's 73-qubit Sycamore [8] and 1121-qubit IBM Condor [9], along with the availability of quantum computing cloud services from providers such as IBM Q Experience [10], Microsoft Azure Quantum [11] and Origin Quantum demonstrate significant advancements in quantum hardware and platforms. This trend in improved hardware utilities is expected to continue.

At the heart of quantum computation lies the quantum circuit, a fundamental construct composed of quantum gates that manipulate qubits, the quantum analog of classical bits. However, harnessing the full potential inherent in quantum circuits presents a formidable challenge. Quantum circuits are highly susceptible to errors and inefficiencies due to quantum noise and the limited capabilities of existing quantum hardware. Present-day quantum processors, which serve as the workhorses for quantum circuit execution, have not yet reached the level of sophistication required for fault-tolerant quantum computing or achieving quantum supremacy. These devices are inherently sensitive to their environment, exhibiting noise, and are susceptible to quantum decoherence. Furthermore, they lack the capacity for continuous quantum error correction. This class of quantum hardware is referred to as Noisy Intermediate Scale Quantum (NISQ) hardware [12]. The confluence of high error rates in the hardware, architectural constraints, limited qubit count, and gate errors due to decoherence collectively stand as a substantial impediment to realizing the full potential of quantum algorithms.

It is well known that quantum computers are highly susceptible to decoherence, which is qubit loss of quantum information over a period of time. Each gate operation introduces a certain amount of noise and can cause qubits to lose their quantum properties. In a large circuit with many gates, errors can propagate through the circuit, which means that if an error occurs early in the circuit, subsequent gates may operate on a corrupted quantum state, leading to a cascade of errors, which can be difficult to mitigate. Hence, the noise level is directly proportional to the quantum circuit size. By reducing the size of the quantum circuit both computational speedup and reduced gate count are achieved. Therefore, the effects of quantum decoherence although not eliminated, are mitigated to some extent. Therefore, optimization by minimizing the number of quantum gates is crucial to the overall reliability and efficiency of quantum computations. Since 2003, when Alfred and Krysta [13] initially proposed addressing the challenge of quantum computing by quantum circuit optimization, this field has evolved into a prominent and actively researched domain in recent years.

Over the past two decades, quantum circuit optimization has undergone significant development. Researchers have devised innovative techniques and algorithms to minimize the number of quantum gates, reduce the quantum circuit's depth, and improve the overall performance of quantum algorithms. These advancements are pivotal in addressing some of the key challenges in the field of quantum computing, such as mitigating quantum hardware constraints and enhancing the practicality of quantum algorithms for real-world problems. This paper broadly discusses circuit optimization based on gate, depth, circuit, and gate fidelity. Gate level optimization refers to reducing the count of quantum gates in the circuit. Depth level optimization refers to increasing parallel computation in the circuit. Circuit level optimization refers to finding the equivalent optimized circuit/sub-circuit. Gate fidelity is the measure of how close the output of a gate is to the expected value of that gate. Optimization techniques to address these challenges may be classified as follows:

1. **Heuristic**: In quantum circuit optimization, this is a method or technique designed to find effective solutions or approximations to a problem using practical and intuitive approaches. The goal is to find the best solution rather than exhaustively searching for all possible ones. Iteratively applying heuristics or rules that improve a circuit's performance can be used to find efficient quantum circuits when used in quantum circuit optimization. In these heuristics, gates are swapped to reduce the circuit depth, adjacent gates are merged to reduce gates, and the gate number is reduced by decomposing multi-qubit gates. Quantum circuit optimization is an NP-hard problem, which makes heuristic algorithms much more practical when applied to large-scale circuits than exact algorithms that promise optimal solutions.
2. **Machine Learning:** Machine learning has shown promising results in several quantum applications, including quantum error correction, quantum control systems, and quantum chemistry simulations. Machine learning model is trained on a set of input-output pairs of quantum circuits, and then the model is used to predict the optimal circuit for a given task.

3. **Unitary Synthesis**: This method focuses on efficiently constructing unitary matrices that represent quantum operations. The factorization process of unitary matrices leads to enhanced performance and a reduction in gate counts, thereby contributing to improved computational efficiency.
4. **Algorithmic approaches:** Where quantum circuit optimization involves systematic methods and procedures designed to enhance the efficiency and performance of quantum circuits. These approaches aim to address specific challenges in quantum computation, such as minimizing gate counts, optimizing resource utilization, and improving overall circuit fidelity.

This paper focuses on the domain of quantum circuit optimization with a primary focus on the methodologies and techniques currently used to enhance the efficiency or speed of quantum computations. This survey will examine various aspects of quantum circuit optimization, including gate-level optimizations, circuit depth reduction, resource-efficient mapping for error mitigation, and their implications for quantum algorithm performance.

The remainder of the paper is structured as follows: Section 2 provides an overview of quantum computing fundamentals, and Section 3 outlines quantum algorithm flows. Section 4 focuses on low-level quantum circuit gate-level optimization classified as level I optimization, and reviews techniques to enhance the efficiency of quantum circuits. Section 5 looks at scale level I optimization of large circuits. Section 6 elaborates on quantum circuit layout optimization, which is classified as level II optimization, emphasizing strategies to improve the spatial arrangement of quantum components. The paper concludes by summarizing key findings and insights drawn from the preceding sections.

## 2. Quantum Computing Basics

The following provides an overview of fundamental concepts in quantum computation that are pertinent to this study. In-depth exploration is given in [14].

In classical computing, the basic unit of information is a bit, which exists in one of two states, 0 or 1. Quantum computing, however, uses quantum bits, or qubits, which can simultaneously represent 0 and 1 due to a phenomenon called superposition. A qubit's state, represented as $|\varphi\rangle$, is a vector in Hilbert space. One qubit is spanned by two possible classical states represented by $|0\rangle$ and $|1\rangle$, where a qubit is a linear combination of $|0\rangle$ and $|1\rangle$, with complex coefficients $a$ and $b$. $|\varphi\rangle = a|0\rangle + b|1\rangle$ where $a, b \in \mathbb{C}$ and $\mathbb{C}$ is the set of complex numbers. The measurement of a qubit collapses its state to either $|0\rangle$ or $|1\rangle$ with probabilities $|a|^2$ and $|b|^2$, respectively.

For a system of $n$ qubits, there are $2^n$ possible basis states. These basis states can be represented with binary values from 0 to $2^{n-1}$. In a classical computer, when you have an $n$-bit input register, it can only exist in one specific state at any given time. It can only be in one state at a time whereas a quantum system in superposition can exist in all possible states simultaneously. This property enables quantum parallelism, allowing a quantum computer to perform a single computation on all possible inputs at once, vastly increasing computational efficiency for specific tasks.

In quantum computing, entanglement is a property that allows qubits to become interconnected, meaning the state of one qubit is dependent on the state of another, regardless of the distance between them. For a two-qubit system (qubits A and B), the Hilbert space is the tensor product of the individual qubit states, with basis states $|0\rangle$A $|0\rangle$B, $|0\rangle$A $|1\rangle$B, $|1\rangle$A, $|0\rangle$B, and $|1\rangle$A $|1\rangle$B, typically written as $|00\rangle$, $|01\rangle$, $|10\rangle$, and $|11\rangle$ Consider a superstition state of two basis states as shown below.

$|\psi\rangle_{AB} = \frac{1}{\sqrt{2}}$ ($|0\rangle$A $|0\rangle$B + $|1\rangle$A $|1\rangle$B represents an entangled state. This specific entangled state is known as a Bell state or EPR (Einstein–Podolsky–Rosen) pair, where the measurement outcome of one qubit A determines the outcome of qubit B, demonstrating perfect correlation. The state $|\psi\rangle_{AB}$ cannot be factorized into the form $|x\rangle_A$ $|y\rangle_B$ for any choice of $|x\rangle_A$ and $|y\rangle_B$. Since $|\psi\rangle_{AB}$ is an entangled state, meaning qubits A and B are not independent; their quantum states are correlated. Thus, $|\psi\rangle_{AB}$ cannot be separated into independent qubit states.

In general, consider a linear combination of *n* qubit quantum states such as $|\chi\rangle = \sum_{i=0}^{N-1} a_i|i\rangle$ ; $N = 2^n$. Accessing information about $|\chi\rangle$ is limited, especially since we cannot directly observe the complex amplitudes $a_i$. When measuring this *n* qubit system in the state $|\chi\rangle$, there are $2^n$ possible outcomes, represented by binary integers from 0 to $2^{n-1}$. A measurement in the standard basis yields the outcome of $|i\rangle^{\text{th}}$ state with a probability equal to $|a_i|^2$.

*Quantum Computing Gates*

Similar to how a classical computer is constructed using an electrical circuit with wires and computations are performed using logic gates, a quantum computer is constructed using a quantum circuit composed of qubit wires and elementary quantum gates. These quantum gates enable the transportation and manipulation of quantum information within the system. A few elementary gates and their unitary representations are shown in Table 1. Quantum computation is reversible, and each quantum gate can be mathematically represented by a unitary matrix, which describes the transformation the gate applies to quantum states. From another perspective, a quantum circuit can be viewed as a composite unitary matrix. This composite matrix can be decomposed into a series of smaller unitary matrices, representing the individual gates applied in the circuit. The computations involved in obtaining a composite unitary matrix for the circuit include matrix multiplications and tensor products (also known as Kronecker products) of smaller unitary matrices.

**Table 1.** Quantum elementary gates.

| Gate Name | # Qubits | Matrix Representation | Mathematical Properties | Quantum Circuit Representation |
|---|---|---|---|---|
| **NOT (X)** | 1 | $\begin{bmatrix} 0 & 1 \\ 1 & 0 \end{bmatrix}$ | $X\lvert 0\rangle = \lvert 1\rangle$<br>$X\lvert 1\rangle = \lvert 0\rangle$ | $\lvert q_0\rangle$ —[X]— |
| **Hadamard** | 1 | $\frac{1}{\sqrt{2}}\begin{bmatrix} 1 & 1 \\ 1 & -1 \end{bmatrix}$ | $H\lvert 0\rangle = \frac{1}{\sqrt{2}}(\lvert 0\rangle + \lvert 1\rangle)$<br>$H\lvert 1\rangle = \frac{1}{\sqrt{2}}(\lvert 0\rangle - \lvert 1\rangle)$ | $\lvert q_0\rangle$ —[H]— |
| **Z-gate** | 1 | $\begin{bmatrix} 1 & 0 \\ 0 & -1 \end{bmatrix}$ | $Z\lvert 0\rangle = \lvert 0\rangle$<br>$Z\lvert 1\rangle = -\lvert 1\rangle$ | $\lvert q_0\rangle$ —[Z]— |
| **S-gate** | 1 | $\begin{bmatrix} 1 & 0 \\ 0 & i \end{bmatrix}$ | $S\lvert 0\rangle = \lvert 0\rangle$<br>$S\lvert 1\rangle = i\lvert 1\rangle$ | $\lvert q_0\rangle$ —[S]— |
| **T-gate** | 1 | $\begin{bmatrix} 1 & 0 \\ 0 & e^{i\pi/4} \end{bmatrix}$ | $T\|0\rangle = \|0\rangle$<br>$T\|1\rangle = e^{i\pi/4}\|1\rangle$ | 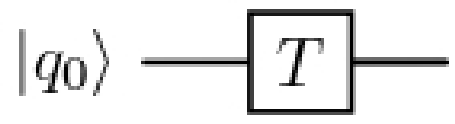 |
| **CNOT(CX)** | 2 | $\begin{bmatrix} 1 & 0 & 0 & 0 \\ 0 & 1 & 0 & 0 \\ 0 & 0 & 0 & 1 \\ 0 & 0 & 1 & 0 \end{bmatrix}$ | $CX\|00\rangle = \|00\rangle$<br>$CX\|01\rangle = \|01\rangle$<br>$CX\|10\rangle = \|11\rangle$<br>$CX\|11\rangle = \|10\rangle$ | 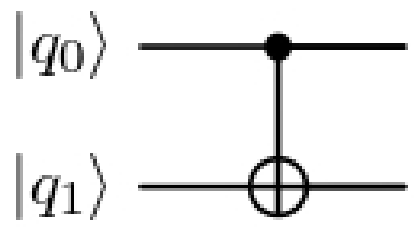 |
| **Controlled Z (CZ)** | 2 | $\begin{bmatrix} 1 & 0 & 0 & 0 \\ 0 & 1 & 0 & 0 \\ 0 & 0 & 1 & 0 \\ 0 & 0 & 0 & -1 \end{bmatrix}$ | $CZ\|00\rangle = \|00\rangle$<br>$CZ\|01\rangle = \|01\rangle$<br>$CZ\|10\rangle = \|10\rangle$<br>$CZ\|11\rangle = -\|11\rangle$ | 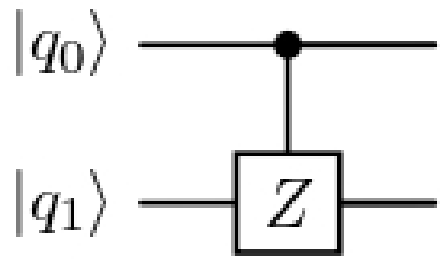 |
| **SWAP** | 2 | $\begin{bmatrix} 1 & 0 & 0 & 0 \\ 0 & 0 & 1 & 0 \\ 0 & 1 & 0 & 0 \\ 0 & 0 & 0 & 1 \end{bmatrix}$ | $Swap\|00\rangle = \|00\rangle$<br>$Swap\|01\rangle = \|10\rangle$<br>$Swap\|10\rangle = \|01\rangle$<br>$Swap\|11\rangle = \|11\rangle$ | 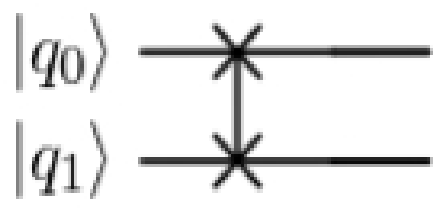 |
| **C-SWAP** | 3 | $\begin{bmatrix} 1 & 0 & 0 & 0 & 0 & 0 & 0 & 0 \\ 0 & 1 & 0 & 0 & 0 & 0 & 0 & 0 \\ 0 & 0 & 1 & 0 & 0 & 0 & 0 & 0 \\ 0 & 0 & 0 & 1 & 0 & 0 & 0 & 0 \\ 0 & 0 & 0 & 0 & 1 & 0 & 0 & 0 \\ 0 & 0 & 0 & 0 & 0 & 1 & 0 & 0 \\ 0 & 0 & 0 & 0 & 0 & 0 & 0 & 1 \\ 0 & 0 & 0 & 0 & 0 & 0 & 1 & 0 \end{bmatrix}$ | $CSwap\|000\rangle = \|000\rangle$<br>$CSwap\|001\rangle = \|001\rangle$<br>$CSwap\|010\rangle = \|010\rangle$<br>$CSwap\|011\rangle = \|011\rangle$<br>$CSwap\|100\rangle = \|100\rangle$<br>$CSwap\|101\rangle = \|110\rangle$<br>$CSwap\|110\rangle = \|101\rangle$<br>$CSwap\|111\rangle = \|111\rangle$ | 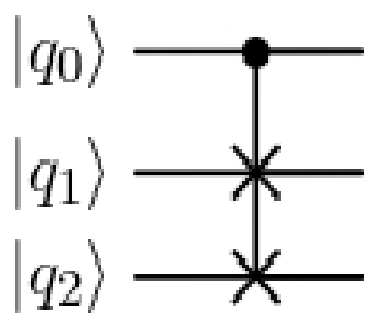 |
| **CCX** | 3 | $\begin{bmatrix} 1 & 0 & 0 & 0 & 0 & 0 & 0 & 0 \\ 0 & 1 & 0 & 0 & 0 & 0 & 0 & 0 \\ 0 & 0 & 1 & 0 & 0 & 0 & 0 & 0 \\ 0 & 0 & 0 & 1 & 0 & 0 & 0 & 0 \\ 0 & 0 & 0 & 0 & 1 & 0 & 0 & 0 \\ 0 & 0 & 0 & 0 & 0 & 1 & 0 & 0 \\ 0 & 0 & 0 & 0 & 0 & 0 & 0 & 1 \\ 0 & 0 & 0 & 0 & 0 & 0 & 1 & 0 \end{bmatrix}$ | $CCX\|000\rangle = \|000\rangle$<br>$CCX\|001\rangle = \|001\rangle$<br>$CCX\|010\rangle = \|010\rangle$<br>$CCX\|011\rangle = \|011\rangle$<br>$CCX\|100\rangle = \|100\rangle$<br>$CCX\|101\rangle = \|101\rangle$<br>$CCX\|110\rangle = \|111\rangle$<br>$CCX\|111\rangle = \|110\rangle$ | 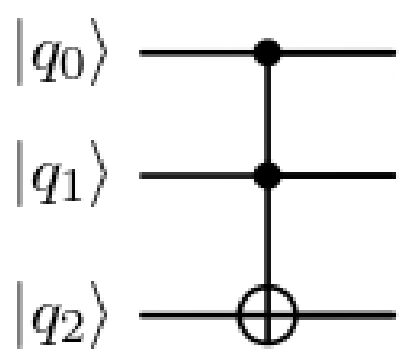 |

## 3. Algorithm Flow in Quantum Computing

Quantum computing can solve problems that were previously considered unsolvable or computationally infeasible for classical computers. Quantum computing takes advantage of the unique principles of quantum mechanics, such as superposition, interference, and entanglement, to achieve a level of parallelism and computational power that surpasses classical computing capabilities [15]. When designing a quantum computing algorithm, the first crucial step is defining a well-structured problem statement. Once the problem is defined, the next step is designing a quantum algorithm that can potentially solve that problem. A quantum algorithm is a set of instructions, or a computational procedure designed to be executed on a quantum computer to solve specific problems more efficiently than classical algorithms. It encapsulates input codes, quantum circuits, and quantum oracles. The quantum oracle is a black box function that is used as input to another algorithm. For instance, in a Grover search, the oracle identifies which values match the search and which values do not match the search. After designing a quantum algorithm for the given problem, a quantum circuit is built. A quantum circuit is a gate representation of a quantum algorithm. It consists of sequences of quantum gates and input qubits. Each gate performs a specific operation on qubits. The circuit represents the sequence of operations required to execute the quantum algorithm and the qubits are initialized and manipulated according to the algorithm's design. Figure 1 illustrates the process of executing a quantum algorithm.

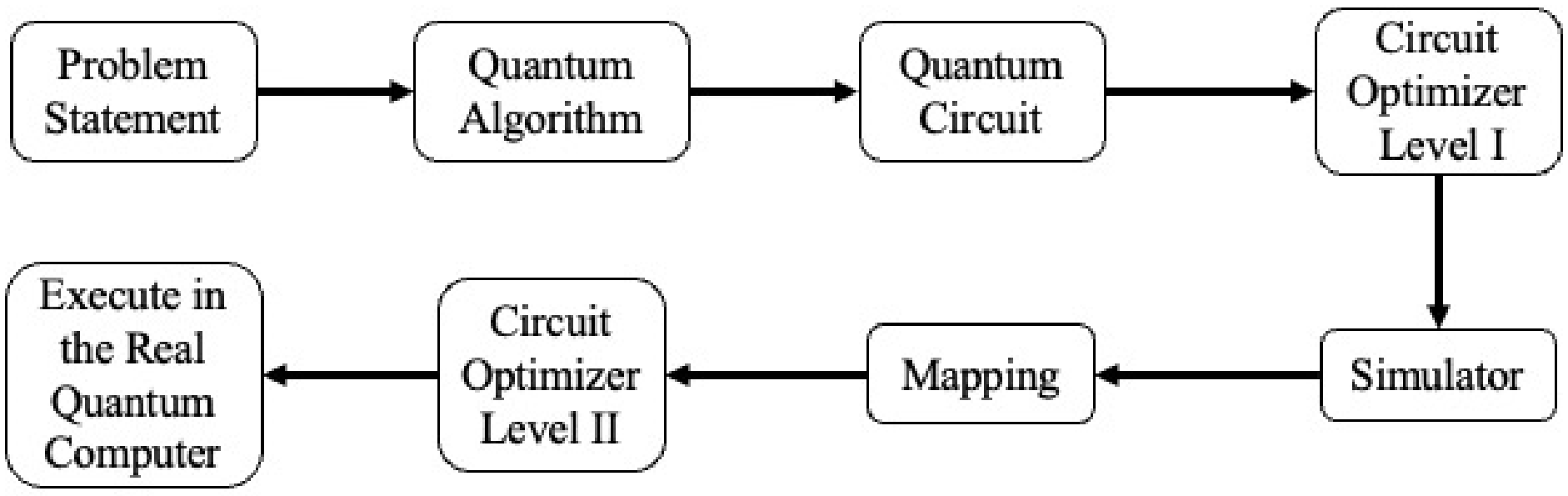

**Figure 1.** Execution pipeline of quantum algorithms.

Depending on the problems and algorithms proposed, quantum circuits can be complex, and circuit optimization helps to leverage computational speed and reliability. Quantum circuit optimization encompasses a suite of strategies, including gate cancellation, gate fusion, and gate reordering, aimed at minimizing the overall circuit depth. The primary objective of these techniques is to decrease both the gate count and depth of quantum circuits, consequently enhancing their efficiency.

Before running the quantum algorithm on a real quantum computer, it is usual to simulate its behavior using a quantum simulator, such as Qiskit [16], Cirq [17], Braket [18], Qutip [19], etc. Quantum simulators are software tools that mimic the behavior of quantum computers on classical hardware. They help to validate the correctness of the quantum algorithm and understand its behaviors without the constraints and noise associated with actual quantum hardware. **Optimization level I** refers to when these optimization procedures are applied prior to simulating a quantum circuit. We will discuss this concept later.

Currently, quantum algorithms are implemented on NISQ architectures. Depending on the hardware platform, there would be a number of constraints, such as geometric limitations (e.g., lack of full connectivity among qubits), quantum gate constraints (e.g., insufficient number of qubits), decoherence limitations (e.g., where qubit loses their information due to external noise or interactions), error rates varying in time and number of qubits. To overcome these limitations and after simulations, the quantum circuit is mapped to the specific hardware environment. This process is referred to as quantum circuit compilation (QCC) or quantum mapping [20]. During the mapping process, the qubit order in the circuit is rearranged according to the underlying hardware to avoid long-time connection. Qubits interact with each other via quantum gates, such as CNOT gates. These interactions are more error-prone if qubits are physically far apart (or have long-time connection) because the error rates increase with the distance over which quantum information must travel.

In physical qubit mapping, qubits that frequently interact are optimally physically placed close to each other, such that decoherence and gate errors are reduced. This is because shorter distances between interacting qubits reduce the time and the noise during operations. To accomplish this task, additional gates may be included to preserve the computation. We will discuss these constraints in more detail in Section 6. Optimization that considers specific hardware qubit mapping constraints and characteristics of the quantum computer is referred to **as optimization level II**. Finally, the optimized quantum circuit or algorithm is run on a quantum computer.

*3.1. Quantum Circuit Optimization*

The complex nature of quantum circuits necessitates meticulous optimization to ensure their efficiency and adaptability for execution on quantum computing hardware. Quantum circuit optimization techniques play a central role in harvesting the full potential of quantum computing, facilitating the efficient solution of real-world problems. In addition to that, circuit optimization helps to mitigate the errors inherent in quantum hardware. In essence, quantum circuit optimization is pivotal in bridging the gap between theoretical quantum algorithms and practical quantum computing, empowering quantum computers to tackle real-world challenges with greater efficiency.

*3.2. Quantum Circuit Optimization Metrics*

Quantum circuit optimization aims to optimize quantum circuits, improve the execution time, require less resource usage, and affect a variety of other factors. In [21], several metrics are used to evaluate the quality and efficiency of optimized quantum circuits. Some of the key metrics used to evaluate quantum circuit optimization are:

*Circuit Depth*: The number of quantum gates executed sequentially in a quantum circuit. Reducing circuit depth is a primary goal of optimization because it can lead to faster execution on quantum hardware.

*Gate Count:* Total number of quantum gates used in a circuit. Minimizing the gate count leads to optimizing resource usage and minimizing the potential for gate errors.

*Qubit Count:* Number of qubits required for a quantum circuit. Reducing the number of qubits is important for optimizing resource usage and for mapping circuits onto quantum hardware with limited qubit availability.

*Two-Qubit Gate Count*: Two-qubit gates are often more resource-intensive and have a higher error probability compared with single-qubit gates. Reducing the number of two-qubit gates in a circuit can significantly impact its efficiency.

*Gate Fidelity:* Gate fidelity measures how closely quantum gates on a quantum computer match the ideal gates. Higher gate fidelity indicates more reliable gate operations, which can impact the overall accuracy of a quantum computation.

*Parallelism:* Parallelism measures the degree to which quantum gates can be executed concurrently in a quantum circuit. Maximizing parallelism can lead to faster quantum computation.

This paper focuses on both optimization level I and optimization level II with two main objectives: 1. Enhancing computational speed through the minimization of resource utilization by specifically focusing on reducing the use of gates and qubits, while also addressing error reduction, called *circuit simplification*. 2. Elevating computational performance by rearranging the order of execution and adjusting qubit mapping while preserving the correctness of the circuit, called *circuit layout optimization*.

## 4. Optimization Approaches

Hardware-independent quantum circuit optimization refers to a set of techniques and methodologies aimed at enhancing the efficiency and performance of quantum circuits without relying on specific details of the underlying quantum hardware. This approach focuses on developing algorithms and strategies that can be applied across different quantum computing architectures, irrespective of their physical implementations. This is classified as Level I Optimization. The emphasis is on overarching methodologies, such as pattern matching and unitary synthesis, which are supported by specific examples of gate cancellations, commutation, and depth reduction. These methodologies provide a unified framework for simplifying quantum circuits and enhancing their efficiency.

### *4.1. Core Methodologies for Circuit Simplification*

**Pattern Matching:** This technique involves recognizing and replacing common substructures in quantum circuits with optimized equivalents. For instance, adjacent gates that cancel or combine into simpler operations are identified and replaced, reducing gate count and depth. Directed Acyclic Graphs (DAGs) are often used to model circuits, representing gates as vertices and dependencies as edges. Gate cancellations and fusions are systematically derived by analyzing the circuit structure.

**Unitary Synthesis:** This strategy focuses on decomposing complex unitary operations into smaller, optimized components. Then, the simplification techniques are applied using machine learning and an algorithmic approach to optimize the circuit.

### *4.2. Simplification Techniques*

This section provides a detailed discussion of circuit simplification rules. These rules are essential for optimizing quantum circuits by reducing the gate count, circuit depth, and resource requirements while preserving the intended functionality of the quantum computation.

#### 4.2.1. Commuting Quantum Gates

Two quantum gates $G_1$, $G_2$ are commuting if, and only if, for any states $|\phi\rangle$, $U_1U_2|\phi\rangle = U_2U_2|\phi\rangle$ holds, where $U_1$ and $U_2$ are unitary matrices corresponding to gates $G_1$ and $G_2$, respectively. $|\phi\rangle$ is also a column vector of the state. Irrespective of the execution orders of gates $G_1$ and, $G_2$, the outcome will be the same. The underlying idea is that rearranging the order of execution while preserving the correctness can increase the computation speed through parallelism as well as find the possibility of any gate cancellations. More than two quantum gates can be commuted; for instance, three gates A, B, and C commute if they satisfy ABC = ACB = BCA = BAC = CAB = CBA. Figure 2 demonstrates basic examples of commuting gates, with additional examples in [21].

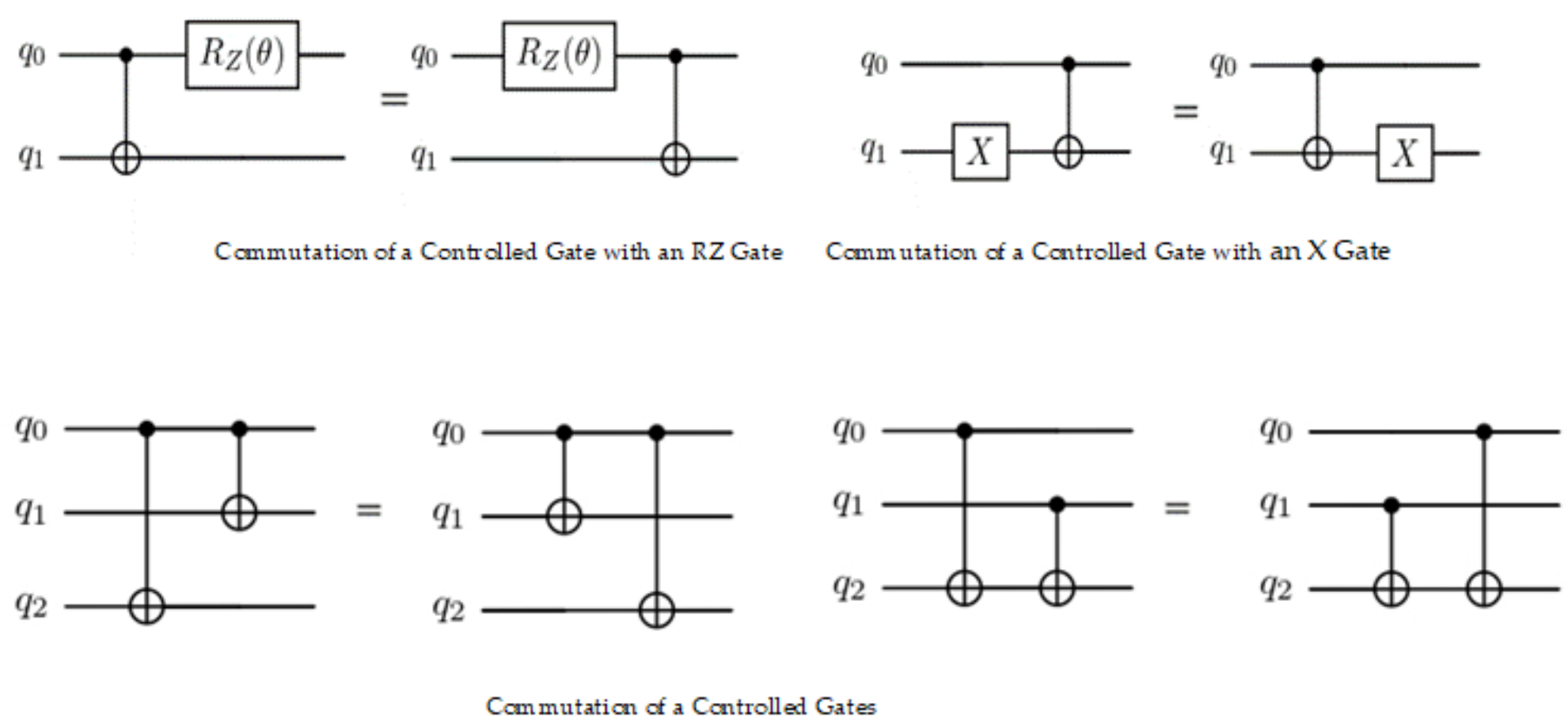

**Figure 2.** Illustration of gate commutations in quantum circuits.

#### 4.2.2. Optimization by Gate Cancellation Rule

Any symmetric matrix U with real entries, is termed a real unitary matrix. All the entries $a_{ij}$ of the matrix are real numbers ($\forall\ a_{ij} \in \mathbb{R}$ ). Then, $\mathbf{U = (U^*)^T}$, where **T** is the transpose of the matrix and * represents the conjugate inverse. If **U** is a real unitary matrix, then it satisfies **U.U = I,** where **I** is the identity matrix. When a unitary matrix multiplies by itself, the result will be unitary. Some examples of real unitary matrices are as follows:

$$\text{NOT gate} = \begin{bmatrix} 0 & 1 \\ 1 & 0 \end{bmatrix}$$

As **X.X = I**, the NOT gate is a real unitary matrix. That is,

$$\begin{bmatrix} 0 & 1 \\ 1 & 0 \end{bmatrix} \cdot \begin{bmatrix} 0 & 1 \\ 1 & 0 \end{bmatrix} = \begin{bmatrix} 1 & 0 \\ 0 & 1 \end{bmatrix} = \text{I}.$$

Other gates, such as the Z gate

$$Z = \begin{bmatrix} 1 & 0 \\ 0 & -1 \end{bmatrix},$$

the Hadamard gate H = $\frac{1}{\sqrt{2}}\begin{bmatrix} 1 & 1 \\ 1 & -1 \end{bmatrix}$, CNOT, SWAP, CCNOT, CSWAP, etc. all have real unitary matrices. The T gate = $\begin{bmatrix} 0 & 1 \\ 1 & e^{i\frac{\pi}{4}} \end{bmatrix}$, and S gate = $\begin{bmatrix} 0 & 1 \\ 1 & i \end{bmatrix}$ have complex numbers in matrix entries ($\forall \alpha_{ij} \notin \mathbb{R}$). Hence, these are not real unitary matrices and **U ≠ (U*)** $^{\mathbf{T}}$ for these gates. In any quantum circuit, when two identical gates are placed next to each other and they have real unitary matrices, the elimination of those gates does not affect the outcome. Refs. [21–25] explore possible gate cancellations for different circuits.

In quantum circuits, adjacent identical gates with real unitary matrices can cancel without altering the circuit's outcome. Refs. [21–23] have examined gate cancellation scenarios for various circuits by using Directed Acyclic Graphs (DAGs) to model quantum circuits, identifying gate dependencies that facilitate cancellations. Similarly, refs. [24,25] use a DAG's canonical form, representing each gate as a vertex and dependencies as edges, allowing analysis of computation flow in the circuit. The DAG framework, applied in quantum circuit optimization, helps visualize gate dependencies, leading to gate cancellation for adjacent, identical gates. Figure 3 shows a quantum circuit C and its corresponding DAG representation.

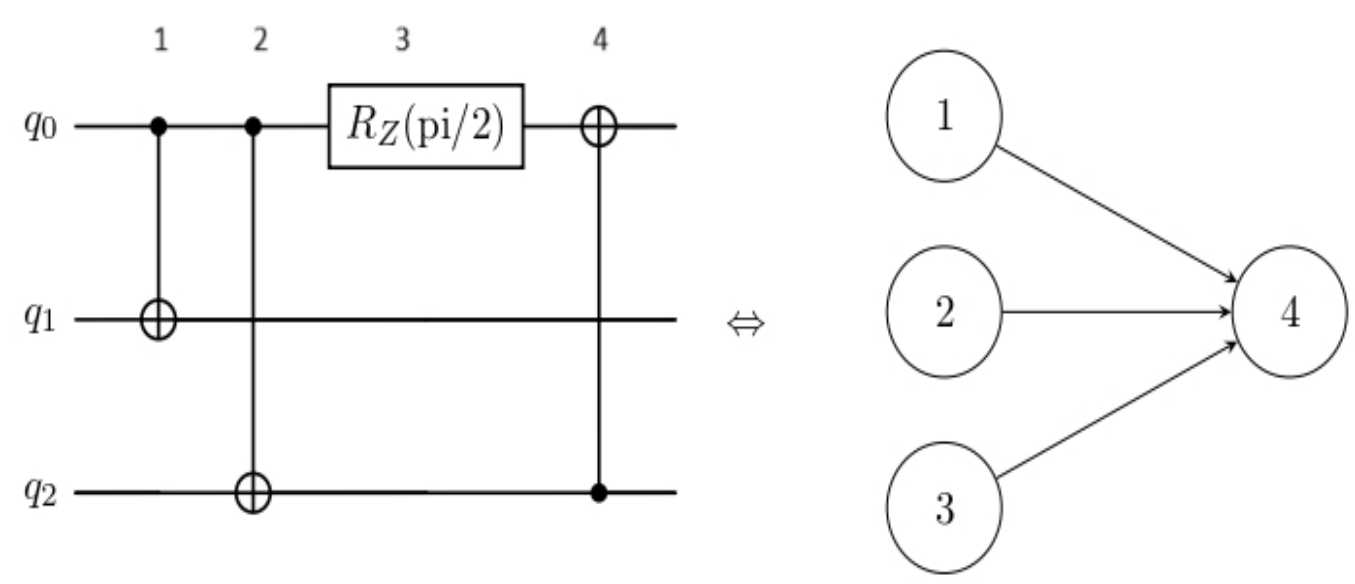

**Figure 3.** Sample circuit C and its associated DAG representation.

In DAG representation, if two adjacent nodes represent the same real unitary gate, they cancel out, yielding the identity matrix. For commuting gates, swapping nodes can lead to cancellations. In certain cases [22], introducing a modification gate enables gate exchange or merging, which can lead to further reductions.

4.2.3. Optimization by Hadamard Gate Reduction

In certain quantum algorithms, the sequence of gates can be optimized by recognizing that you only need specific combinations of Clifford gates to achieve your desired transformation. By simplifying the gate sequence, you can reduce the number of Hadamard (H) gates required. Figure 4 illustrates a few examples where the Hadamard gate count is reduced. Refs. [21–25] discuss more Hadamard reduction scenarios.

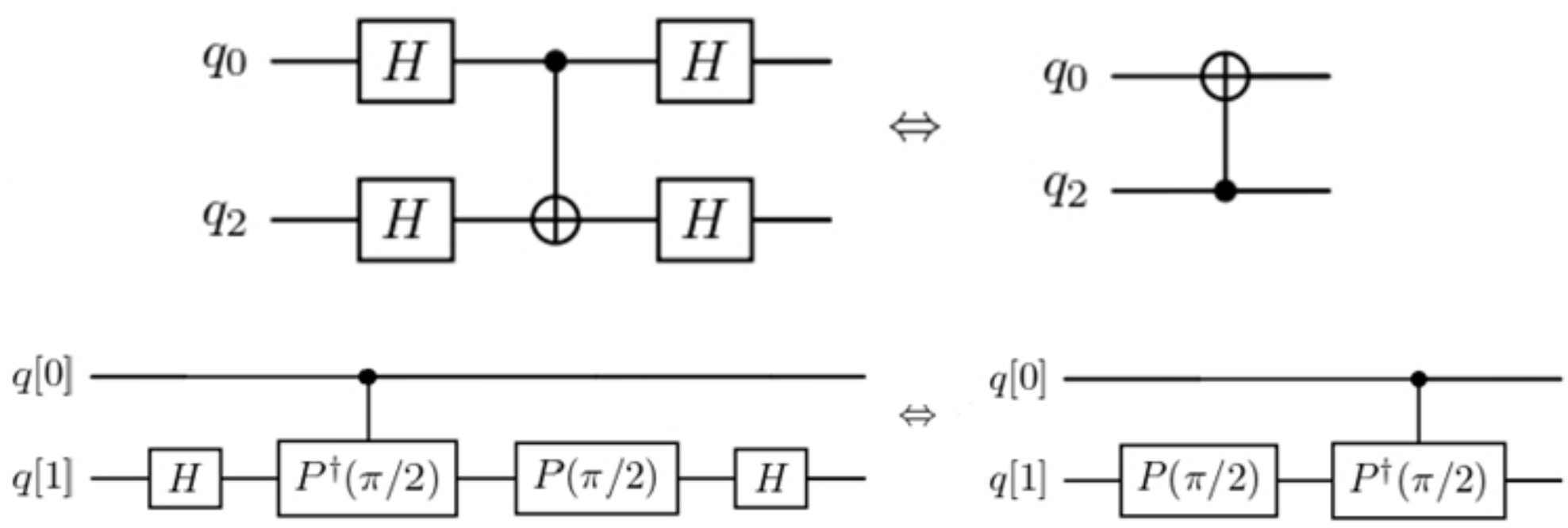

**Figure 4.** Illustrations of Hadamard gate reduction.

In [25], the authors propose a comprehensive 3-level strategy aimed at optimizing circuits through reversible mapping and quantum levels to reduce the H-count. At the reversible level, the logic gates and their configurations are optimized to ensure minimal quantum gate usage and prepare for effective quantum-level mapping. The mapping level focuses on the efficient allocation of quantum gates to physical qubits in the quantum processor. This step is crucial to minimize latency and the overall gate count, further enhancing the circuit's performance. At the quantum level, the operational parameters of the quantum gates are fine-tuned, and their coherence and control are optimized to achieve the best possible H-depth outcomes. H-depth refers to the depth of a quantum circuit when only considering the Hadamard (H) gates.

#### 4.2.4. Merging Rz Gates Using Phase Polynomial Estimation

This strategy focuses on decomposing complex unitary operations into smaller, optimized components. Specifically, phase polynomial estimation is employed to merge rotation gates (Rz gates) when the circuit comprises only CNOT, NOT, and Rz gates. This approach leverages phase dependencies, effectively reducing both the circuit depth and gate count, as proposed in [26].

Imagine adjusting multiple radio dials to tune into specific frequencies. If two dials are set to the same frequency, you do not need both—you can combine them into one dial with the sum of their adjustments. Similarly, in a quantum circuit, when two rotation gates share the same pattern, they can be merged into a single gate with a combined rotation angle. In other words,

- Every RZ gate contributes a rotation angle to the quantum state.
- Each gate's contribution depends on how the input qubits interact with the rotation operation.
- These interactions follow specific patterns or dependencies based on the input state.

If two rotation gates produce the same phase pattern, they can be merged into one gate with a single, combined rotation angle. This merging reduces the number of gates without changing the circuit's final behavior.

### 4.3. Depth Reduction

The depth of a quantum circuit is defined as the number of sequential layers or time steps required to execute a given quantum computation. It is a crucial parameter in measuring the overall performance of a quantum algorithm. So far, we have been discussing how to reduce computations by gate cancellations and gate fusions. Rearranging the order of gates, while preserving the correctness of the algorithm at the same time, can also lead to minimizing the number of computations.

In [27], Zhu et al., proposed depth optimization for linear reversible circuits. Circuits which implement the linear building blocks are called linear reversible circuits, which only consist of CNOT gates. e.g., stabilizer circuits. Any CNOT gate controlled by the *j*th qubit, acting on the *i*th qubit ($i \neq j$), can be written a $E_{ij} = I + e_{ij}$ . Where *I* is the identity matrix and $e_{ij}$ is a matrix with all entries equal to 0, except for entry (*i, j*), which is one. It is well known that any invertible matrix can be decomposed into a combination of elementary matrices. $U = P \prod_{i=k}^{1} E(c_i, t_i)$, where *P* is a permutation matrix.

Consider the decomposition sequence SEQ of a unitary operator U as a finite sequence of elementary matrices with a particular order. SEQ = {$E(c_1, t_2)$, $E(c_2, t_4)$, $E(c_3, t_1)$, …, $E(c_k, t_k)$}, where $E(c_i, t_i)$ is the elementary matrix representation of a CNOT with control qubit $c_i$, target as qubit $t_i$. The elementary matrix/operator (CNOT) can be commutable if $\cap_i = \{c_i, t_j\} = \emptyset$ for any $i$ and $j$ in SEQ, which means for any two adjacent gates $E(c_k, t_k)$ and $E(c_{k+1}, t_{k+1})$ in SEQ, the order of the two gates can be swapped if, and only if, $c_k \neq t_{k+1}$ and $c_{k+1} \neq t_k$.

For example, in Figure 5, the given quantum circuit (Top left) has a depth of 7. SEQ = {E (0, 1), E (1, 3), E (2,3), E (2, 0), E (0, 3), E (0, 2), E (2, 1)}. According to the rule described above, the adjacent gates E(1,3)and E(2,3) can be swapped since their control and target qubits do not overlap in a way that creates a dependency, meaning the control qubit of one gate is not the target qubit of the other ( $c_k \neq t_{k+1}$ and $c_{k+1} \neq t_k$ ). Thus, the depth becomes 4 as shown in Figure 5 (bottom).

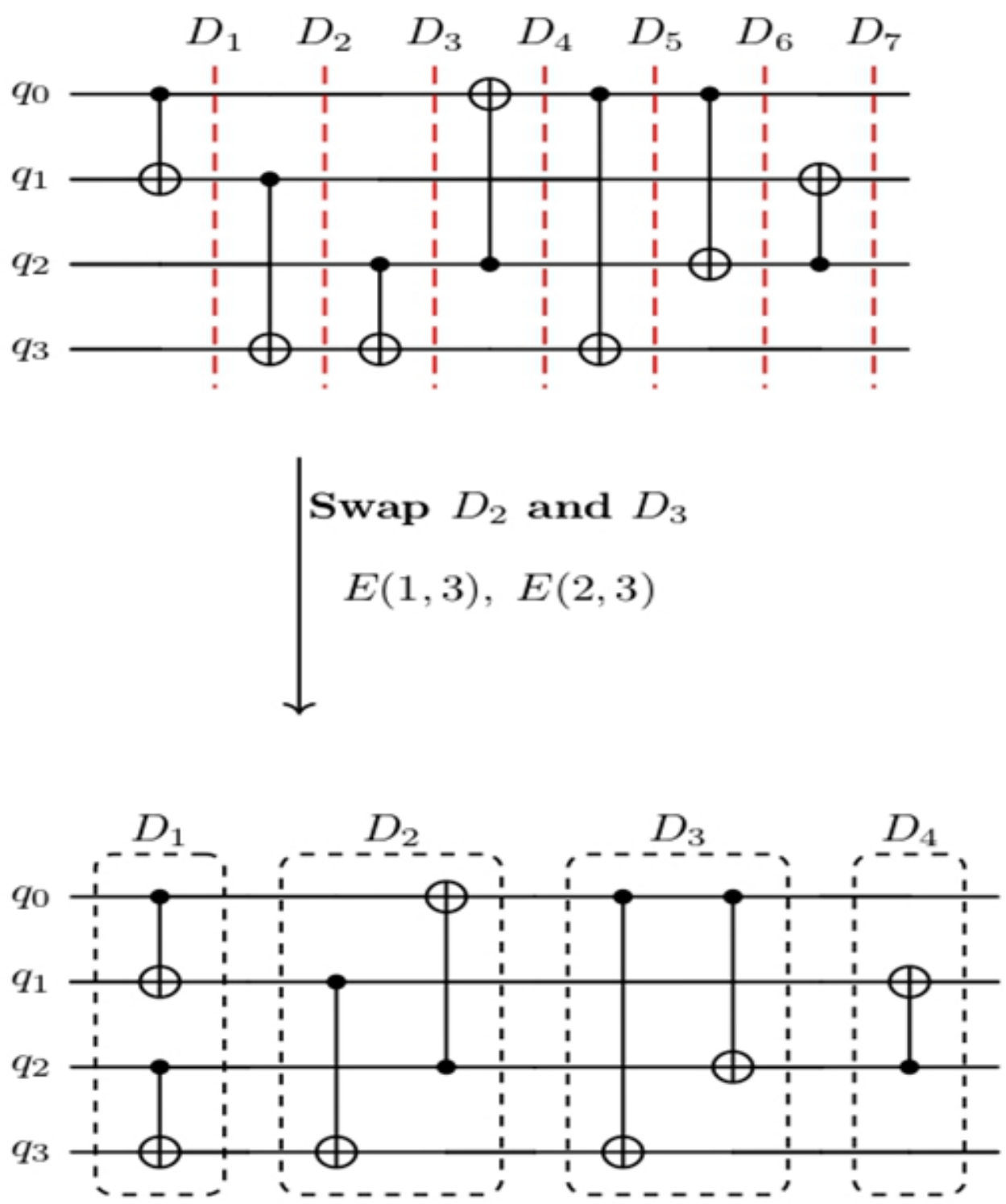

**Figure 5.** Illustration of depth reduction by linear arrangements.

In [28], proposed that depth reduction is achieved primarily through the parallel execution of gates by leveraging commutation relationships within the cost operator, which often consists of independent terms acting on subsets of qubits. This allows gates to be executed simultaneously, reducing sequential layers. However, the circuit depth grows linearly with the number of QAOA layers (p), which limits scalability for near-term devices. The method's effectiveness depends heavily on the problem structure; for dense graphs or highly connected cost operators, fewer gates can commute. In [29], Babbush et al. found sparsity in the Hamiltonian allows gates corresponding to non-interacting terms to be executed in parallel. In fermionic systems (fermionic systems refer to physical systems that are governed by fermions, which are particles that obey the Pauli exclusion principle. Examples of fermions include electrons, protons, and neutrons), the Hamiltonian often contains local terms that act on disjoint qubit pairs, and because these terms commute (i.e., they do not interfere with each other), the corresponding operations can be applied simultaneously. This parallel execution reduces the number of sequential steps in the quantum circuit. Additionally, techniques like Givens rotations and fermionic swap networks are employed to reorder gates efficiently and to further decrease them. But in systems with dense interactions, where many terms overlap, fewer operations can be parallelized, limiting the benefits. In [30], Quantum Fourier Transform (QFT) circuits, controlled-phase gates, are applied sequentially, leading to the risk of depth increment. By leveraging ancilla qubits, these gates can be precomputed and stored, allowing phase shifts to be applied simultaneously across multiple qubits. This eliminates sequential dependencies and reduces the depth of controlled-phase operations. Furthermore, the recursive structure of QFT benefits from ancilla qubits, as the circuit can be broken into smaller, independent subproblems that execute in parallel. The overall circuit depth is thereby reduced from $O(n^2)$ to $O(n \log n)$. In [31], Draper et al. use a similar technique to store carry in quantum arithmetic operations to reduce the depth. Ref. [9] utilizes this technique to optimize modular arithmetic operations in Shor's algorithm, improving depth efficiency by leveraging ancilla qubits for intermediate computations and carry propagation. However, this improvement comes at the cost of increased qubit usage, as more resources are required to manage the ancilla qubits. More qubits increase susceptibility to noise and decoherence, particularly in NISQ devices.

In [32], Kun Zhang proposed methods to reduce the circuit depth of Grover's algorithm, addressing its limitations for near-term quantum devices. The authors introduce a critical ratio, the ratio between the depths of the oracle and the global diffusion operator, as a threshold parameter to determine when the global diffusion operator can be replaced with a local diffusion operator. The global diffusion operator performs a reflection of all quantum states around their mean amplitude, requiring operations across the entire Hilbert space, which leads to significant circuit depth. In contrast, the local diffusion operator performs a similar reflection but acts only on specific, smaller subsets of the state space, reducing computational dependencies and enabling parallel execution of gates. By identifying when the critical ratio is less than one, the authors demonstrate that replacing the global operator with local versions significantly reduces depth while maintaining Grover's quadratic query complexity.

The replacement of the global diffusion operator with the local one not only reduces depth but also allows for further optimization. It enables commutation between consecutive oracle operators, creating opportunities for gate cancellation and gate fusion (combining multiple gates into single operations). Furthermore, the authors introduce a multi-stage quantum search strategy, where the search process is recursively divided into smaller, rescaled subsets of the database. This approach reduces the overall search space and enhances depth efficiency without increasing the number of oracle calls. The proposed optimizations maximize parallelism, making the algorithm more practical for near-term quantum hardware with limited coherence times. However, the local diffusion operator reduces depth; it introduces additional complexity in identifying and implementing the appropriate local operations, requiring extra pre-processing overhead.

In summary, depth reduction is essential for improving the performance of quantum circuits, particularly for near-term devices. Techniques such as gate rearrangement, ancilla-assisted optimizations, and specific algorithmic strategies (e.g., for QAOA or Hamiltonian simulation) play a vital role in achieving shallower, more efficient circuits. This example of optimizing CNOT sequences demonstrates how careful analysis, and reordering can significantly reduce circuit depth while preserving correctness

### *4.4. Challenges*

While these optimization techniques yield significant benefits, they come with challenges, including the following: Techniques like ancilla-assisted optimizations require additional qubits, which may not be feasible for near-term devices. Advanced optimizations, such as unitary synthesis or pattern matching, introduce computational overhead for identifying patterns and dependencies; Noise Sensitivity. Reducing depth often increases qubit usage, leading to higher susceptibility to noise and decoherence on NISQ devices.

## 5. At-Scale Optimization

Implementation of optimization level I at a large scale, by using the heuristic simplification rules discussed in the previous section, such as gate fusion, gate cancellation, and rearranging the execution order for gate/depth reductions, can be extended to automate quantum circuit simplifications effectively. Three primary methodologies are widely adopted for large-scale optimization:

1. Artificial intelligence (AI)-based simplification
2. Synthesizing unitary matrices
3. Algorithmic approaches

These methodologies address different aspects of circuit simplification and are summarized below, with detailed explanations.

### *5.1. Optimization Using Artificial Intelligence-Based Approaches*

Since the early 1990s, AI has become an essential tool in optimization, offering flexible and innovative ways to solve complex problems in a variety of fields [33]. The main advantage of automation in optimization is that it can handle large, complex quantum circuits that would be difficult to process using traditional optimization methods. AI algorithms can learn from data to make predictions and decisions about optimizing a given objective function [33]. These techniques can adapt to changes in the optimization problem, such as changes in the input data or constraints, without requiring manual updates to the optimization strategy. In this section, we discuss machine learning approaches currently being used in quantum circuit optimization.

#### 5.1.1. Optimization Using Reinforcement Learning

Reinforcement learning (RL) [34–37] is a type of machine learning model that focuses on training agents to learn from experience and make optimal decisions in each environment. RL has been applied to a wide range of problems, including game playing, robotics, and natural language processing.

In RL, an “agent” interacts with the rest of the world or the “environment”. In several iterations, the agent receives information from the environment and, in response to this observation, chooses an action which alters the state of the environment. The agent adapts its strategy to maximize a success measure, the “reward”. Figure 6 illustrates a reinforcement learning process where an agent iteratively optimizes a quantum circuit. At time t1, the agent extracts a subcircuit from a larger circuit, referred to as the environment At. The agent then searches for the best match from the Q-table. Based on this match, the agent takes actions AE(St+1) to modify the circuit. It receives feedback in the form of a reward Rt+1, which is used to refine its actions over time, leading to continuous improvement.

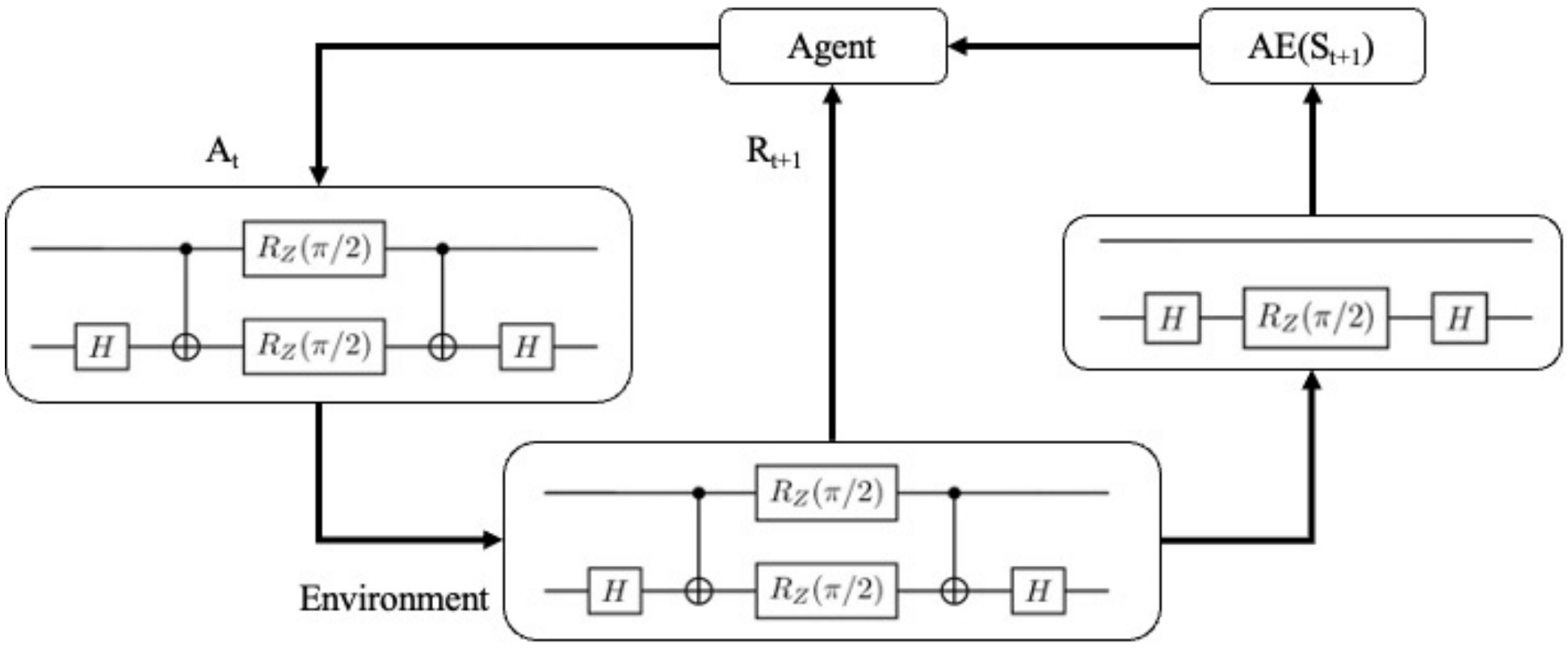

**Figure 6.** Flow diagram for quantum circuit optimization using RL.

In [38], Fosel et al. propose a novel reinforcement learning (RL) framework to automate this process, enabling the systematic reduction of gate counts and circuit depths. This method leverages the inherent flexibility of RL to identify and apply a sequence of circuit transformations, improving performance while preserving the intended functionality of the circuit.

At the heart of the framework are two types of transformation rules. **Soft rules** involve local modifications, such as gate fusion and rearrangement, which reorganize the sequence of gates for improved efficiency. On the other hand, **hard rules** focus on removing consecutive quantum gates that cancel each other, leading to a direct reduction in gate count. Each transformation is applied independently at different locations in the circuit, and the RL agent learns to prioritize these rules through a trial-and-error process, guided by a reward policy that incentivizes effective optimization.

A key innovation of this approach is its representation of quantum circuits in a 3D grid format, designed to capture the intricate structure of quantum operations. The three axes of the grid represent the quantum circuit index, which identifies the circuit being optimized; the moment **(timestamp)**, which corresponds to the parallel execution of gates; and the gate class, which categorizes gates. Each timestamp is treated as a subcircuit, encapsulating gates that can be executed simultaneously. This structured representation allows the RL agent to systematically explore equivalent circuits within the action space, searching for transformations that minimize circuit complexity.

The RL framework relies on a robust mapping of states and actions. In traditional reinforcement learning, this mapping is stored in a Q-Table, which grows as the agent explores new configurations. The Q-Table initially starts small but expands rapidly during the exploratory phase, before stabilizing as the agent begins to exploit its accumulated knowledge. Each entry in the Q-Table represents a subcircuit and its equivalent optimized form, encoding the agent’s experience. However, in deep reinforcement learning, the Q-Table is replaced by a machine learning model that dynamically learns the mapping between states and actions. This enhancement allows the model to scale more effectively, handling larger circuits and more complex optimization tasks.

The optimization process begins with the RL agent interacting with the circuit environment. At each step, the agent evaluates potential transformations, applying either soft or hard rules to improve the circuit. Based on a predefined reward policy, the agent selects the transformation that offers the greatest benefit, iteratively refining its strategy. Over time, the agent develops a comprehensive understanding of the optimization space, identifying the most efficient configurations for the circuit. The final optimized circuit is then decoded from the 3D representation, yielding a design that is both functionally equivalent and resource efficient.

In conclusion, the reinforcement learning framework introduced by Fosel et al. represents a significant advancement in quantum circuit optimization. By combining robust circuit representation, flexible transformation rules, and dynamic learning, this method provides a powerful tool for improving the efficiency of quantum circuits. As quantum computing continues to evolve, such automated optimization techniques will play an essential role in bridging the gap between theoretical designs and practical implementations. Although the RL-based approach to quantum circuit optimization is a promising step forward, its practical implementation is hindered by computational costs, scaling issues, and sensitivity to noise and hardware constraints.

In [39], the authors propose a reinforcement learning (RL) framework to optimize the design and architecture of variational quantum circuits (VQCs). The methodology leverages a policy-based RL agent to iteratively refine VQC structures by dynamically selecting gates, qubits, and circuit layouts to maximize performance. The agent explores the space of possible circuit architectures and is trained using reward functions based on the quality of the quantum output, such as energy estimation accuracy for tasks like solving the Variational Quantum Eigensolver (VQE) problem. The approach is distinct in its focus on optimizing the circuit architecture itself, rather than just parameter tuning, which allows for more efficient and tailored circuit designs for specific tasks. However, the exploration phase can be computationally expensive, as evaluating the performance of each candidate circuit requires multiple quantum simulations or hardware executions. Additionally, the reliance on predefined reward functions and training data limits the method's ability to generalize to entirely new quantum tasks or architectures.

Moro et al. [40] used a deep reinforcement learning (DRL) framework to optimize quantum circuit compilation. The methodology employs an RL agent to decompose target unitary operations into sequences of quantum gates tailored to specific hardware constraints. The agent interacts with a simulated environment representing the quantum circuit and learns to generate optimal gate sequences through trial and error. The framework incorporates a policy network that predicts gate sequences and a value network to estimate future rewards, trained using Proximal Policy Optimization (PPO). The reward function incentivizes the agent to minimize the cost of the compiled circuit, balancing gate count, depth, and fidelity. This approach demonstrates significant improvements in scalability and efficiency over traditional heuristic-based compilation methods, especially for complex quantum circuits. However, the framework has some limitations. Training the DRL model requires extensive computational resources and simulation time, making it challenging for larger systems.

Similarly, Salehi et al. [41] propose an optimization framework aimed at improving the efficiency of quantum circuits used in quantum machine learning (QML). The methodology involves a rule-based approach where gate elimination and circuit simplifications are applied iteratively to reduce the resource utilization (e.g., gate count and depth) while maintaining the performance of the quantum model. The framework leverages heuristic rules to identify redundancies and symmetries within the circuit, enabling transformations that preserve the functionality of the QML model. The optimization process also incorporates a fidelity measure to ensure that the modified circuits closely approximate the original computational outcomes. This method is particularly advantageous for QML tasks, as it minimizes computational overhead and improves scalability for larger datasets. However, the approach has limitations. The reliance on heuristic rules means the optimization process may not be universally applicable to all QML architectures, particularly those with unconventional or highly entangled structures.

In [42], Riu et al. introduced a reinforcement learning (RL) framework for quantum circuit optimization that leverages the ZX-calculus, a graphical language for quantum mechanics. The methodology employs Graph Neural Networks (GNNs) to represent and process the intricate dependencies within quantum circuits, effectively capturing their graph-like structures. The RL agent, trained using the Proximal Policy Optimization (PPO) algorithm, applies ZX-calculus simplification rules to iteratively reduce circuit complexity. This approach enables the agent to identify and execute optimizations that minimize gate counts and circuit depth, enhancing overall efficiency. However, the computational intensity associated with training the GNN-based RL agent is significant, particularly for large-scale circuits. The complexity of the ZX-calculus rules and the vast action space can lead to extended training times and increased resource consumption. Additionally, the effectiveness of the optimization heavily depends on the quality and comprehensiveness of the training data, which may limit the agent's performance when encountering novel or highly complex circuit configurations.

#### 5.1.2. Optimization Generative Adversarial Network

Ref. [43] proposed a Quantum Generative Adversarial Network (QuGAN) framework to generate efficient quantum circuit approximations with a focus on minimizing the gate count and depth while maintaining high fidelity. The methodology leverages quantum state fidelity as a training metric, ensuring that the generated circuits closely approximate the target quantum states. The QuGAN architecture consists of a quantum generator and a quantum discriminator, both of which are iteratively trained; the generator creates candidate circuits, and the discriminator evaluates their fidelity against the target states. By optimizing the fidelity measure, the framework identifies circuit configurations with reduced complexity. This approach demonstrates the potential for resource-efficient quantum circuit synthesis, particularly for applications like state preparation in quantum chemistry. However, a key drawback of QuGAN is its reliance on fidelity-based training, which becomes computationally expensive for large-scale systems due to the need for repeated state evaluations.

### *5.2. Optimization by Synthesizing Unitary Matrices*

Another method for automating quantum circuit simplifications is by synthesizing unitary matrices. This approach involves using mathematical techniques to analyze and manipulate quantum circuits at a more fundamental level by decomposing quantum gates into their underlying unitary matrices [44–46]. Techniques like matrix factorization and decomposition algorithms help to find simpler representations of quantum gates or find equivalent sequences of gates that achieve the same quantum operation but with fewer resources.

Quanto, which is the first quantum circuit optimizer that automatically generates circuit identities (equivalent optimized circuit) is proposed in [47]. Quanto can optimize a wide range of quantum circuits, including circuits with non-Clifford gates, multi-qubit gates, and controlled gates. To demonstrate Quanto's effectiveness, the authors have used it to optimize quantum circuits, such as quantum error correction circuits, surface code circuits, and quantum circuits for solving optimization problems. In all cases, Quanto significantly reduced the number of gates and improved the overall performance of the circuits. Pointing et al. [45] developed a novel approach to optimizing quantum circuits by exploiting circuit symmetries. For the development of practical quantum computers, the Quanto technique shows promising results and can enhance the performance and scalability of quantum circuits significantly Quanto proposes compiling a quantum circuit to an optimized executable circuit for target hardware using synthesis. Quantum circuit synthesis is defined by decomposing a quantum circuit into smaller sequences of unitary matrices. This approach proposes a hierarchal step to optimize the given quantum circuit. It has been proven that more than 30% of CNOT gates are reduced by this method.

In [48], Mingkuan Xu et al. (2022) present an automated optimization framework for quantum circuits that combines equivalence checking, super optimization, and backtracking techniques. Quartz first generates Equivalence Circuit Classes (ECCs) by iteratively identifying alternative representations of subcircuits, allowing for transformations such as gate cancellations, merging Rz gates, and depth reduction. Using equivalence rules and pruning redundant operations, Quartz employs a backtracking search to systematically explore possible optimizations, ensuring global circuit minimization. The optimized circuits are then converted into symbolic representations for compatibility with hardware-specific transformations. This methodology achieves significant gate and depth reductions, particularly for circuits with non-Clifford gates or high connectivity constraints. However, a notable drawback is the computational overhead of the equivalence-checking and backtracking process, which can scale poorly for large circuits. Additionally, the reliance on pre-defined equivalence rules limits adaptability to novel circuit structures or hardware-specific optimizations. Despite these challenges, Quartz demonstrates substantial improvements in circuit efficiency, particularly for NISQ-era devices.

In [49], Wu et al. propose an optimization framework that leverages automated synthesis techniques to reduce the depth and gate count of quantum circuits while maintaining scalability for large systems. QGo employs a divide-and-conquer strategy, breaking down circuits into smaller subcircuits that can be optimized independently using rewrite rules and equivalence transformations. By automating the identification and application of these optimizations, QGo integrates hardware-specific constraints, ensuring compatibility with quantum processors. The framework iteratively refines circuits by exploring alternative configurations by balancing the circuit complexity with execution efficiency. QGo achieves scalability by partitioning large circuits and optimizing them in parallel, leveraging a dynamic programming approach to merge subcircuit optimizations. However, its reliance on heuristic-based synthesis can lead to suboptimal results for certain circuit configurations, as the framework may overlook deeper global optimizations. Additionally, the partitioning process can introduce overhead and dependency constraints, limiting the efficiency of parallel optimization in circuits with high connectivity.

 According to Shende, Markov, and Bullock [50], universal n-qubit circuits must contain at least $\frac{1}{4}(4^n - 3n - 1)$ CNOT gates. Michal Sedlák et al., ref. [51], propose an approach to find, for an arbitrary given unitary operator, a quantum circuit containing the lowest possible number of CNOT gates using the singular value decomposition (SVD) method.

In [52], Zhang et al. showed, by utilizing the properties of decomposition of diagonal invertible matrices which can be commutable, that any quantum diagonal unitary operator can be decomposed as Rz and CNOT gates. Commuting the execution order of the gates helps to cancel the gates and possible parallelism on execution which results in depth optimization.

In [53], Markov and Shi (2008) outline procedures that leverage tensor network representations to enhance the efficiency of quantum circuit simulations, with implications for circuit optimization. The key strategy involves representing quantum operations and states as tensors, enabling sparse representations that reduce computational overhead by avoiding redundant operations. By optimizing the order of tensor contraction, the computational cost of simulating large circuits is minimized, which parallels gate reordering and depth reduction in circuit synthesis. The authors further exploit the structure of quantum circuits by decomposing large tensors into smaller, manageable substructures, facilitating parallel execution and resource efficiency. Graph-theoretic methods, such as minimizing the treewidth of the tensor network graph, ensure that intermediate tensor sizes are kept minimal, correlating to reduced ancillary resources and depth in circuits. Additionally, handling high-dimensional systems via factorization of unitary matrices allows the identification and removal of redundant components, directly translating to gate reduction. While the focus is on simulation, these techniques provide a robust framework for optimizing quantum circuits, to reduce gate count, depth, and resource utilization. Despite their strengths, while tensor networks effectively reduce computational overhead for many quantum circuits, their efficiency depends on the structure and sparsity of the circuit. For highly entangled systems or dense circuits, the tensor contraction process can become computationally expensive due to the rapid growth of intermediate tensor sizes. For circuits with complex connectivity or large treewidth, the optimization benefits diminish, and the contraction order becomes computationally intensive to determine. Identifying the optimal contraction order and decomposing tensors requires significant preprocessing. This overhead can be a bottleneck.

The methodologies discussed in this section highlight the significant advancements in quantum circuit synthesis and optimization, focusing on techniques to simplify circuits, reduce gate counts, and minimize depth. Approaches such as automated synthesis frameworks (e.g., Quanto, Quartz, and QGo) and mathematical decomposition methods (e.g., SVD and tensor networks) provide robust tools for tackling the challenges of resource-intensive quantum circuits. These techniques demonstrate remarkable efficiency in reducing gates like CNOTs, merging rotation gates, and leveraging equivalence rules for global circuit minimization. Furthermore, methods like tensor network representations and hierarchical decompositions enable scalable solutions for high-dimensional systems, while incorporating hardware-specific constraints ensures compatibility with current quantum devices. However, these approaches are not without limitations, including computational overheads, dependency on pre-defined equivalence rules, and challenges in scalability for highly entangled or densely connected circuits. Despite these drawbacks, the continuous evolution of quantum circuit synthesis techniques, particularly for NISQ-era devices, underscores their critical role in advancing the practical implementation of quantum algorithms and enhancing the overall performance of quantum systems.

### *5.3. Optimization Using Algorithmic Approaches*

An algorithmic approach [54,55] involves the systematic and methodical application of algorithms, which are sets of precise instructions and procedures, to solve complex problems, streamline processes, or accomplish specific goals. By breaking down intricate tasks into manageable steps, this structured methodology ensures efficiency and accuracy in problem-solving, decision-making, and data processing.

Quantum circuit optimization can be automated by variational algorithms. In [56–58], the underlying principle for quantum classical fusion is the variational principle. This principle can be compared to a radio tuner. The quantum circuit is analogous to the electronic radio circuit. The quantum circuit consists of an encoding layer and a trainable layer. The encoding layer consists of rotation gates where the angles of the rotation gates become the parameters that can be tuned by the classical component. This parametrized circuit is called the ansatz. The classical component is analogous to the tuner/capacitor in the radio. The tuner is used to tune in order to match the frequency of the radio station. Likewise, the classical layer is used to tune the parameters of the quantum circuit until the cost function associated with the quantum circuit is optimized. The Variational Quantum Eigen solver reduces the quantum resources or quantum gates that would be required in a quantum circuit.

Ref. [59] proposes Variational Gate Circuit Optimization (VQGO). VQGO uses a Variational Quantum Eigensolver (VQE). The multi-qubit quantum circuit is initially parametrized into single-qubit gates and source gates. This decomposition into elementary gates is essential as the input multi-qubit gate circuit may not be compatible with running on specific quantum hardware, whereas elementary gates, such as source gates or single qubit, would be available on the hardware. VQGO uses these single-qubit gates with parametrized rotation angles along with a drive to create multi-qubit gates. The optimization is performed by iterating through the rotation angles of the single-qubit gates that are based on the cost function's measured values. Figure 7 illustrates the VQGO method. The cost function is the Average Gate Infidelity (AGI). AGI compares two quantum channels and checks how far on average they are from the pure states.

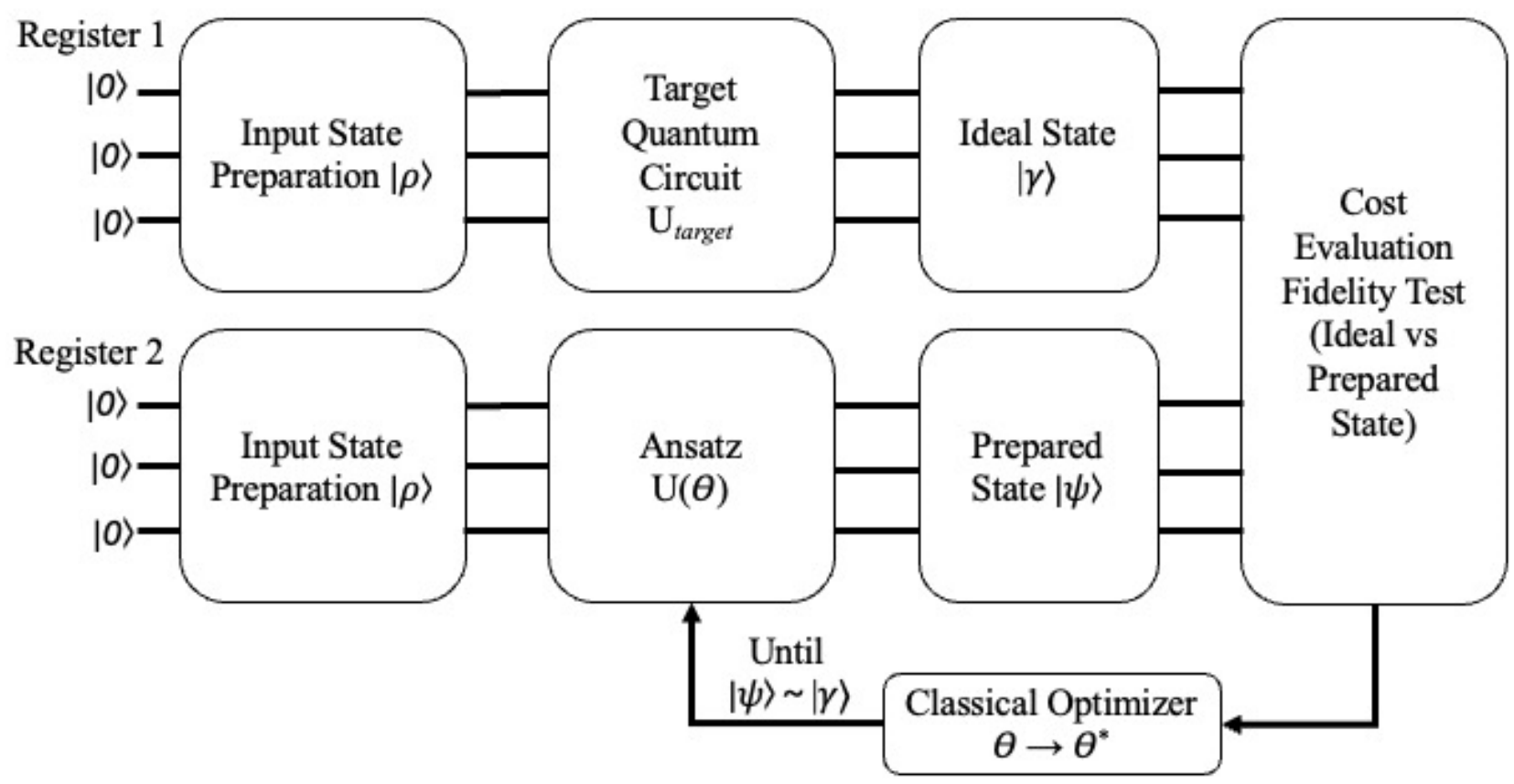

**Figure 7.** Illustration of hybrid quantum circuit optimization.

The cost function depends on the target circuit $U_{\text{target}}$ parameters and the ansatz circuit $U(\theta_l)$ parameters. The objective is to align the target circuit to the ansatz circuit. Initially, a high-cost function would be obtained because of the difference between the two quantum circuits. With every iteration, the parameters of the target circuit are adjusted to reflect the pure state parameters. The target circuit is optimized when the AGI value associated with the two quantum channels is minimal. VQGO can achieve a higher fidelity compared with conventional methods even in cross-resonance environments. Cross-resonance is interference or unwanted interaction between qubits, which can lead to errors in quantum operations. In a CNOT gate, cross-resonance can result in the control qubit interacting with other qubits in the circuit, which results in the target qubit flipping or changing its state even when it was not intended. The comparative table, Table 2, highlights the distinctions between classical machine learning (ML) approaches and hybrid quantum-classical methods for quantum circuit optimization.

**Table 2.** Comparison between classical optimization model with hybrid approach.

| Feature | Classical Optimization Models | Hybrid Quantum-Classical Optimization Models |
|---|---|---|
| **Core Principle** | Fully reliant on classical computational resources and algorithms. | Combine quantum computing for specific tasks with classical optimization layers. |
| **Problem-Solving Scope** | Suitable for well-defined, small to medium-sized optimization problems. | Designed for problems where quantum advantages can be leveraged (e.g., VQE, QAOA). |
| **Algorithm Examples** | RL, GNN, GAN | Variational Quantum Eigensolver (VQE), Quantum Approximate Optimization Algorithm (QAOA). |
| **Scalability** | Efficient for many large-scale problems, though constrained by classical limits. | Scale better for specific quantum problems but limited by quantum hardware capabilities. |
| **Optimization Speed** | Fast due to mature classical computing technologies. | Slower due to iterative communication between quantum and classical layers. |
| **Cost Function Dependency** | Classical cost functions with no quantum measurements. | Hybrid cost functions often incorporate quantum measurements for updates. |

In [60], gate-free implementations of the encoding circuit of VQEs are also shown. Control VQEs showed significant improvement in the speed of the state preparation handled by hardware-controlled pulses as compared to the encoding circuits of VQEs. Stenger et al. [61] proposed a pulse-scaling method that translated the area of the CR and rotary pulses to the 3-dimensional space with RzX(θ) rotations. Their approach shows an increase in gate fidelity without any added calibrations. The above work is extended into even arbitrary gates to construct a circuit transpiration framework that decomposes two-qubit gates into the RzX rather than the traditional C-NOT transpiration.

In [62–65], genetic algorithms (GAs) mimic nature's process; that is, individuals better adapted to the environment have a greater chance of surviving and passing on their genetic traits to their offspring. In the context of GAs, a candidate solution is represented as a chromosome, and a group of these solutions forms a population. The encoding of chromosomes can vary depending on the specific problem and it may utilize binary strings, direct value representations, and tree structures. Ref. [66] proposed a genetic algorithmic method for optimizing quantum circuits. Circuit optimization is formalized as follows: N quantum circuits form the initial population. Each circuit's fitness is determined by its output state vector. From the population, one circuit is selected and a cross-over made with the one that needs to be optimized. The crossover would be swapping a single time step or multiple steps of the circuit. The mutation is also performed by flipping a single-qubit gate, interchanging the control and target of control gates, and tuning the parameter for the rotation gates for exploring new circuits. The newly generated circuit is added to the populations depending on the fitness score. The above process is repeated until the desired optimum is reached. In [67], Wei et al. generated the population by using a tree-structured method, where they incorporated the equivalent circuit for each leaf as well as each subtree. This helps to reduce the time complexity of the algorithm. Peng et al., in [68], successfully optimized the quantum teleportation circuit using a genetic algorithm.

The methodologies discussed here, provide a comprehensive framework for large-scale quantum circuit optimization. Table 3 describes the comparative analysis in detail. AI-based approaches excel in adaptability and iterative learning, while unitary synthesis enables resource-efficient transformations through mathematical techniques. Algorithmic approaches, on the other hand, offer systematic strategies tailored for specific goals, such as depth reduction or fidelity improvement. Each method has distinct advantages and drawbacks, making their combination a promising direction for further research in NISQ-era quantum computing.

**Table 3.** Comparative analysis of optimization methodologies.

| Methodology | Key Techniques | Advantages | Drawbacks |
|---|---|---|---|
| **AI-Based Approaches** | Reinforcement Learning, Deep RL, GAN | Adaptive, scalable for complex circuits | High computational demands |
| **Unitary Synthesis** | Matrix Decomposition | Gate and depth reductions | Computational overhead, matrix structure dependency |
| **Algorithmic Approaches** | Variational Algorithms, Genetic Algorithms | Hardware-aware, systematic optimization | Time-intensive, computational complexity |

## 6. Quantum Circuit Layout Optimization

Quantum circuit layout optimization is a level II optimization. The practical realization of theoretical quantum circuits on quantum hardware is a complex process [69,70]. Each hardware has specific physical layouts of qubits. In a superconducting qubit computer, every qubit in the layout cannot be connected with every other qubit. But theoretical quantum computational models assume that direct interactions between any two physical qubits are always possible. However, in actual physical implementations, establishing direct interactions between distinct qubits can be extremely challenging—sometimes, it may not be possible. For example, in the IBM QX4 architecture [12] shown in Figure 8, $Q_1$ can directly interact with $Q_0$ only. $Q_2$ can directly interact with $Q_0$ and $Q_1$. There is no direct interaction possible between $Q_3$ and $Q_1$, $Q_0$ to $Q_1$, and so on.

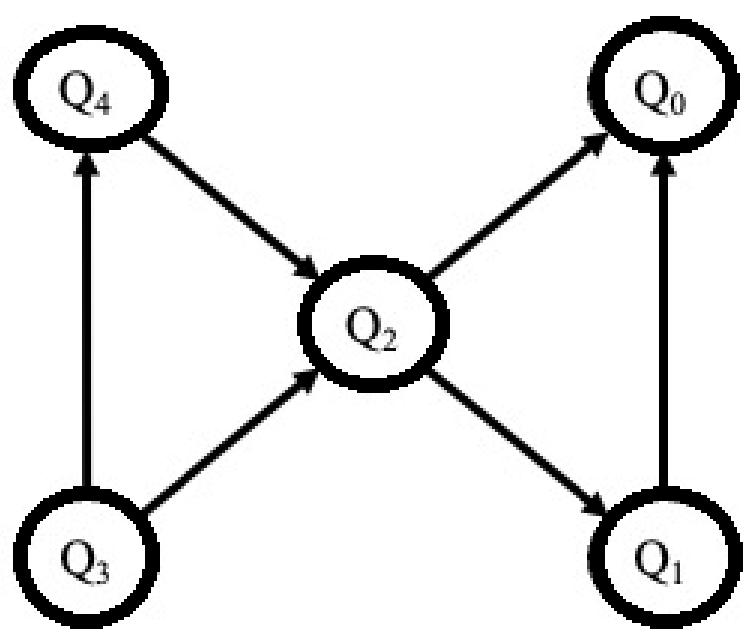

**Figure 8.** IBM QX4 architecture.

In addition, certain qubit pairs may have interaction frequencies as low as 0.2 Hz; for example, carbon qubit pairs C2 and C6 have interaction frequencies, while decoherence processes can take around one second. This means that the interaction between such qubits would generate undesirable noise. The connectivity between qubits determines how efficiently these algorithms can be implemented. If qubits are not directly connected, additional operations may be needed to enable their interaction, which can increase the complexity and runtime of the algorithm. The fewer additional gate operations a quantum computation needs, the faster it can execute and the less chance there is for errors to creep in. The connectivity varies greatly between various quantum hardware.

Mapping the simulated circuit to a specific hardware is a crucial step in the process of implementing an algorithm on a quantum computer. It becomes necessary to introduce additional bridge gates, such as SWAP, H, and CNOT gates, into the theoretical or simulated circuits. This mapping process is called quantum circuit placement (QCP). It incurs overhead in terms of computational resources and time and harms the fidelity of the execution. To address this issue, several approaches have been proposed in the literature to minimize or optimize these bridge gates.

Quantum circuit placement ensures that the logical qubits of an algorithm are mapped efficiently onto the available physical qubits while adhering to the device's connectivity constraints. Figure 9 shows an example of QCP. The initial circuit (Figure 9a) has multiple layers of gates applied across four qubits. For instance, Hadamard (H) gates and CNOT gates are placed based on the logical operations required by the algorithm and mapped directly to the available qubits. The optimized circuit in Figure 9b shows fewer gates and a more compact arrangement. For instance, CNOT gates are now placed in a way that reduces the need for intermediary swaps, and Hadamard gates are applied more efficiently. The qubits $q_3$, $q_2$, $q_0$, and $q_1$ are reordered for more efficient gate count reductions.

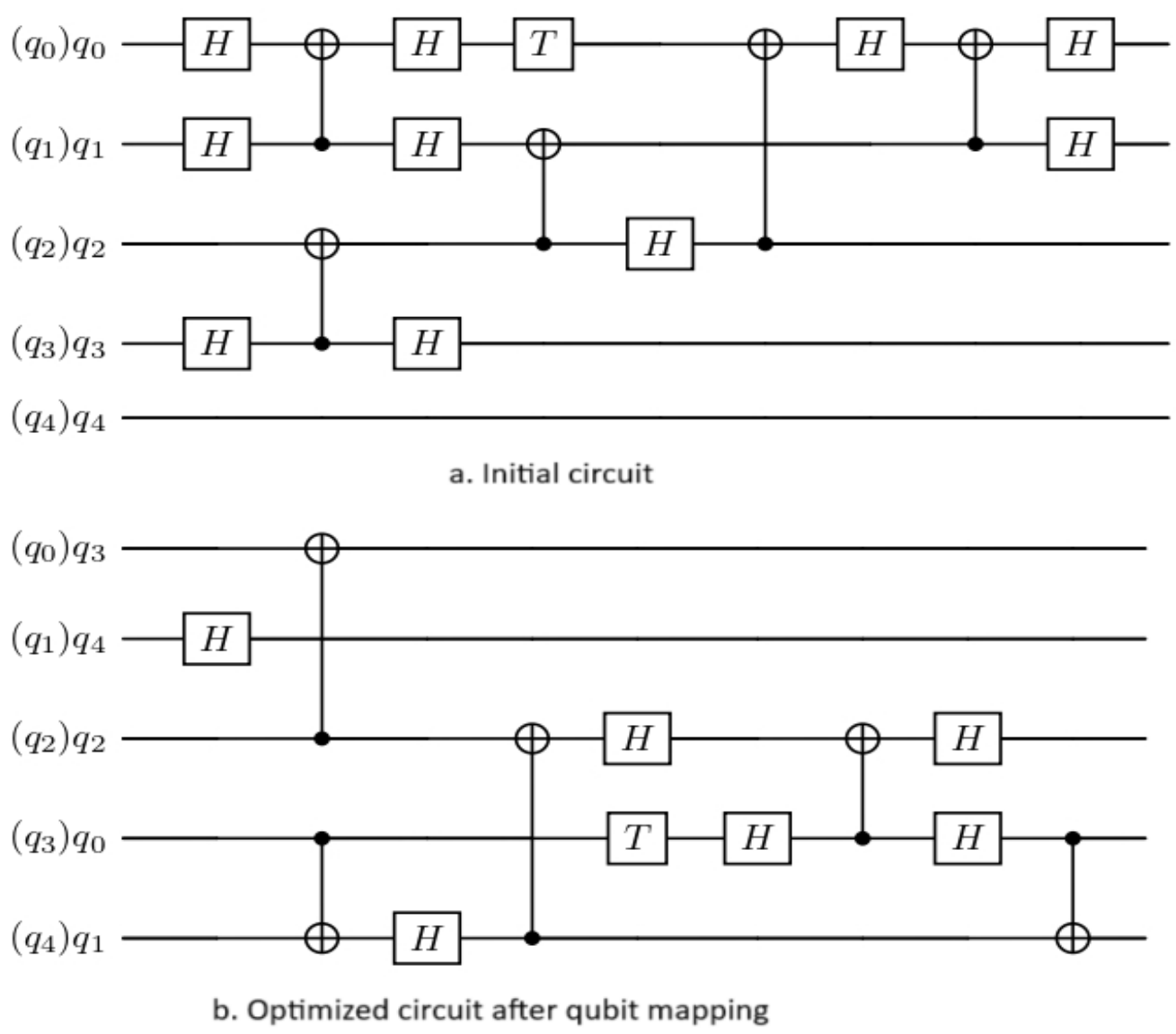

**Figure 9.** Quantum circuit placement illustration.

Optimizations might include merging gates, eliminating redundant gates, and reducing gate operations on each qubit to reduce error rates and execution time on quantum hardware. Layout optimization is an NP hard problem, which preserves the theoretical circuit functionality while aiming to increase computational speed. This is called **optimization level II,** which can be defined as optimization by adding additional bridge gates to satisfy the mapping constraints; that is, the mapping of logical or simulated circuits to a physical layout. In this section of the paper, we will briefly delve into the current trends and approaches aimed at tackling this intricate problem.

Ref. [71] formulates the QCP as a search problem and successfully demonstrates the work of on two IBMQ 20-qubit systems, named Tokyo and Poughkeepsie with fidelity as the benchmark. The graph structure of quantum circuits is used to predict their expected fidelities for certain devices considering their respective noise characteristics. An idea related to the one proposed in this work, which follows a more coarse-grained approach, is demonstrated in [72], where a supervised machine learning model is used to predict good combinations of devices and existing compilation flows. RL is applied to learn a strategy for applying individual gate transformation rules to optimize quantum circuits.

Ref. [73] proposes avoiding long interactions by considering the linear nearest neighbor computational architecture where all qubits are lined up in a chain and only the nearest neighbors are allowed to interact. Mathematically, for a system with n qubits $q_1, q_2, \dots q_n,$ $q_1$, the 2-qubit gates are allowed on qubits whose subscript values differ by one. However, this is not necessarily possible in real quantum circuits. Any circuit with gates spanning more than two qubits is translated into a circuit with a single qubit, and two qubit gates. Next, permutations are generated of possible connections between qubits using graph theory, where each qubit can be represented as a node in a graph, the connections between them can be depicted as edges, and the coupling strength between qubits as the weight of the edges. Using a dynamic programming approach the optimum topology map is selected for the required circuit. Wille et. al. formulates the QCP as a symbolic optimization problem [74]. Only multi-qubit gates are impacted by connectivity constraints [75]. Hence, before mapping the logical circuit into the physical qubits, the coupling map of a device is checked, and the optimum mapping solution is chosen using a Boolean satisfiability solver to minimize the Hadamard (H) and SWAP operations on each timestamp.

Bhattacharjee et. al. [76] used a multi-tier method to address the QCP problem, which includes the selection of topology, resolving swap gate constraints, and choosing the QC map which yields the highest fidelity. Itoko et al. [77] proposed an algorithmic approach for compiler optimization using gate transformation and commutation to reduce swap and bridge gates. Besides circuit optimization at the logical level, using the assertion method, Haner et al. [78] show that entanglement description can also improve circuit optimization for a specific target architecture.

Fan et. al. [79] proposed an algorithm that entails training a reinforcement learning (RL) agent to discover an optimal swap strategy for randomly generated mappings. The agent’s objective is to select the best strategy from a pool of possible combinations, where each combination consists of a random mapping paired with a specific swap strategy. They formulate QCP as a bi-level optimization problem. At level I, the framework finds the optimum placement mapping for a given quantum circuit. At level II, the framework focuses on reducing the swap gate cost. To achieve this, they use a deep learning RL model. The input to the model is a state matrix S and the initial mapping of qubits. The $i$-th row of the state matrix represents the $i^{th}$ physical qubit and its associated logical qubit. Each column of the matrix state represents a separate time step (level). The element in the $i^{th}$ row and the $j$th column denotes the position of the qubit connected to the $i^{th}$ qubit, which is specified by its corresponding quantum gate at time step $j$. The value is set to be −2 if no quantum gates are associated with the qubit at a certain time step. The underlying algorithm uses all the SWAP insertions allowed by the target quantum devices as the action space. The reward policy is described as follows: (i) gate reward, which equals the number of quantum gates executed given the current action, the updated state, and hardware constraints; (ii) done reward, which is applied when all the quantum gates have been executed; (iii) SWAP penalty, when a SWAP gate is inserted; and (iv) non-execution penalty, when there are no executable gates after the SWAP gate is inserted.

In [80], Paler et al. proposed a mapping technique that helps to reduce bridge gates. They experimented with both heuristic and machine learning approaches. They formulated the mapping problem as a weighted random search (WRS) problem. The optimum mapping is selected by assigning weights to parameters such as the maximum depth of the tree (number of bridge gates required), the maximum number of children (qubits), and the configuration of the search space. After that, the algorithm is run iteratively to find the best mapping techniques based on reducing the overall weight as much as possible. This approach is called QXX. Once the initial solution is obtained, this will be used as learning data for the machine learning approach QXX-MLP; next, the parameters weights are tuned to improve the QXX solutions. Figure 10 illustrates the process of the Paler et.al mapping method.

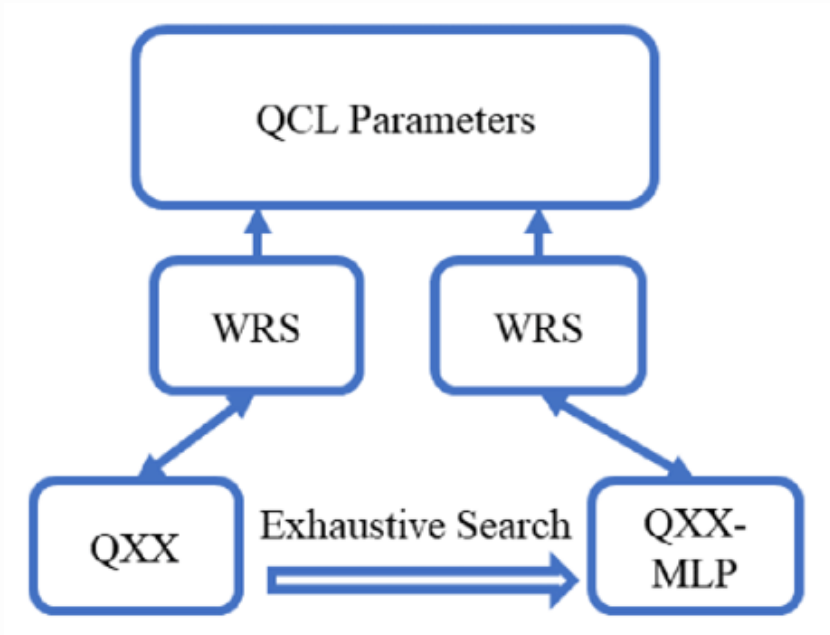

**Figure 10.** Continuous learning QCP.

Ref. [81] presents a methodology integrating reinforcement learning (RL) to optimize quantum circuit synthesis and routing. In circuit synthesis, the task is modeled as a sequential decision-making process where an RL agent iteratively selects gates to transform an initial operator into the identity operator, effectively constructing the desired circuit. During training, the agent follows a curriculum learning approach, gradually tackling more complex target operators, and is rewarded for successful synthesis while penalized for excessive gate usage to encourage efficiency. After training, the agent synthesizes circuits by applying its learned policy, ensuring near-optimal solutions that align with device-specific constraints. Similarly, in circuit routing, RL is employed to optimize SWAP operations, which adapt circuits to hardware coupling maps. The agent evaluates potential SWAPs to minimize the qubit distances for forthcoming two-qubit operations, reducing the circuit's depth and gate count. Together, these RL-based approaches streamline circuit optimization for diverse quantum hardware environments. However, challenges related to scalability, training overhead, adaptability, and reward function design need to be addressed to fully realize its potential in practical quantum computing workflows.

In [82], Siraichi et al. (2018) address the challenge of mapping logical qubits in quantum programs to physical qubits on hardware with restricted connectivity. They propose a qubit allocation framework that optimizes the placement of logical qubits to minimize the need for additional SWAP gates which are required to satisfy connectivity constraints. The methodology involves modeling the allocation problem as a graph optimization task, where the connectivity of physical qubits is represented as a graph, and logical qubit operations are mapped to nodes. The framework uses a cost model to evaluate different mappings based on metrics such as the gate fidelity, latency, and the SWAP gate overhead, selecting the optimal allocation through search algorithms. Techniques like lookahead heuristics and backtracking further refine the mapping to reduce runtime overhead. While effective, the framework has drawbacks: the optimization process can become computationally expensive for circuits with high qubit counts or dense connectivity requirements, limiting scalability. Additionally, the reliance on static cost models may not fully account for dynamic noise and hardware variability, potentially leading to suboptimal allocations for specific devices.

In [83], Zulehner et al. propose an optimization framework to map quantum circuits onto IBM QX devices which have limited qubit connectivity. The methodology involves transforming the logical qubit interactions in a quantum circuit to conform to the hardware's physical constraints while minimizing the insertion of SWAP gates. The mapping process is formulated as a search problem, where the authors employ **A*** search algorithms to explore optimal solutions by considering both the immediate and future costs of qubit placements. A key innovation is the use of lookahead mechanisms to evaluate the long-term impact of SWAP gate insertions, leading to more globally optimized solutions. Additionally, the framework introduces heuristics for dynamically selecting SWAP operations that minimize the gate count and circuit depth. While this approach significantly improves the efficiency of mapping circuits to IBM QX architectures, its main drawback is the computational overhead associated with the A* search, which scales poorly for large circuits. Furthermore, the reliance on static heuristics can lead to suboptimal performance when hardware noise and variability are not accounted for.

Ref. [84] introduced a novel approach to compiler optimization known as "Relaxed Peephole Optimization" (RPO). This approach operates under the assumption that certain qubit states in quantum gates can be determined during compilation. Leveraging this assumption, RPO offers the advantage of replacing quantum operations with more cost-effective structures (such as gates or layers) while preserving their equivalence. One notable advantage of this approach, in contrast to earlier techniques as documented, is its inherent capacity to dynamically alter the unitary matrix during runtime without affecting the outcome. In their work, the authors introduced a relaxed peephole optimization for specific quantum gates and well-defined input states like basis and pure states.

To distinguish between quantum states, the author introduced two optimization passes, namely, the Quantum Basis State Optimization Pass and the Quantum Pure State Optimization Pass. In their paper, they defined $|0\rangle$, $|1\rangle$, $|+\rangle$, $|-\rangle$, $|L\rangle$, and $|R\rangle$ as basis states.

$$|+\rangle = \frac{|0\rangle+|1\rangle}{\sqrt{2}}\ ;\ |-\rangle = \frac{|0\rangle-|1\rangle}{\sqrt{2}}\ ;|L\rangle = \frac{|0\rangle+i|1\rangle}{\sqrt{2}}\ ;\ |R\rangle = \frac{|0\rangle-i|1\rangle}{\sqrt{2}}$$

In each of the passes, a partial state automaton was performed to show how they can analyze the inputs (quantum gates) acting on specific states (basis or pure states). The implementation of the Quantum Basis State Optimization Pass (QBO) and the Quantum Pure State Optimization Pass (QPO) was carried out within the open-source quantum computing framework Qiskit.

Ref. [85] demonstrated an automated method to recompile a quantum circuit A into a target circuit B, with the aim of achieving identical results for a specific input, as expressed by the equation B |in> = A |in>. The process is initiated with the user defining B as an empty template consisting of parameterized unitary gates set to the identity matrix. Subsequently, the compilation process is carried out utilizing quantum hardware to execute an isomorphic energy-minimization task. An isomorphic energy-minimization task refers to the process of finding an alternative representation or layout of a quantum circuit while minimizing the energy or cost associated with the quantum gates and operations used in the circuit. To overcome potential slow convergence in variational eigen-solving techniques, the approach explored involves using a sequence of intermediate targets that act as "lures" to guide and accelerate the optimization process toward the final target. Below is a comparative analysis in Table 4. Summarizing content in this section.

**Table 4.** Comparative table for layout optimization.

| Methodology | Key Innovations | Advantages | Drawbacks |
| --- | --- | --- | --- |
| **Bhattacharjee et al. [76]** | Multi-tier method involving topology selection and swap constraint resolution. | High fidelity through tiered optimization. | Computationally expensive for dense connectivity requirements. |
| **Itoko et al. [77]** | Algorithmic gate transformation and commutation optimization to reduce SWAP and bridge gates. | Logical-level optimization and improved circuit efficiency. | Limited focus on physical hardware constraints. |
| **Fan et al. [79]** | RL-based framework using deep learning for QCP as a bi-level optimization problem. | Adaptive learning of SWAP strategies; effective handling of dynamic mapping problems. | Training overhead; scalability issues; challenges in reward function design. |
| **Paler et al. (QXX and QXX-MLP) [80]** | Weighted random search for mapping optimization and machine learning-based parameter tuning. | Iterative improvement of mapping techniques; reduced overall gate weight. | Limited scalability for large circuits. |
| **Siraichi et al. [82]** | Graph optimization task for qubit allocation with cost models and lookahead heuristics. | Minimized SWAP gate overhead, improved fidelity and latency. | Computationally expensive for high qubit counts; static cost models limit adaptability to dynamic noise. |
| **Zulehner et al. [83]** | Formulated QCP as a search problem using A* search algorithms and lookahead mechanisms to minimize SWAP gates. | Globally optimized solutions with reduced gate count and circuit depth. | Computational overhead; poor scalability for large circuits; static heuristics may not handle hardware variability effectively. |
| **Relaxed Peephole Optimization (RPO) [84]** | Replacing quantum operations dynamically during runtime. | Cost-effective gate replacements with runtime adaptability. | Applicability limited to specific input states and gates. |
| **Automated Circuit Recompilation [85]** | Energy minimization task for recompiling circuit A into target circuit B using intermediate targets. | Accelerated convergence towards optimized circuit representation. | Potential for slow convergence in certain cases; dependency on intermediate “lure” targets. |

## 7. Review and Future Directions

While quantum computing represents a revolutionary technological advancement, the development of real-time quantum applications remains a big challenge. This challenge is primarily attributed to the Noisy Intermediate-Scale Quantum (NISQ) nature of current quantum devices, which are inherently susceptible to noise and errors. Though there are ongoing efforts to improve error correction and mitigation technologies, quantum circuit optimization is one of the promising approaches to do so. These techniques are instrumental in economically mitigating errors and enhancing the overall performance and reliability of quantum computations.

The phase polynomial method discussed in [21] encounters limitations when applied to quantum circuits restricted to certain gate sets, such as CNOT, NOT, and Rz gates. Among these, the Hadamard (H) gate holds particular significance in quantum computing, as it facilitates the creation of superposition states and entanglement, which are foundational to quantum algorithms. Quantum circuits can employ a wide range of gates, including SWAP, CSWAP, T, and P gates, each contributing uniquely to the circuit’s overall functionality. Efficiently merging Rz gates with these different gate types is crucial for optimizing circuit performance. However, as the number of gates to be merged increases, challenges arise in maintaining the accuracy of phase polynomial estimations.

The phase polynomial approach mathematically represents the circuit's operations and interactions, particularly focusing on Rz gates. When dealing with many gates, inaccuracies in this representation can lead to suboptimal gate merging. Such inaccuracies can adversely affect the circuit's overall computational efficiency and fidelity, potentially undermining its ability to execute tasks with precision.

Optimization based on reinforcement learning (RL) considers both physical and logical level quantum circuit optimization achieved by learning from past experiences and optimizing itself over time. Ref. [37] demonstrates optimizing certain types of quantum gates, such as X, Z, and CNOT gates. To generalize this algorithm for all elementary gates by adding more circuit patterns to help the agent learn, it may be able to generalize its knowledge to new circuits with similar patterns. However, adding additional circuit patterns to the training data can also cause a problem called Qtable exploitation. This happens when the agent focuses on many specific models of the training data and does not explore other possible solutions.

In summary, quantum circuit optimization begins with encoding the quantum circuit into a structured representation, such as tuples, tree structures, symbolic forms, or graphs. The optimization problem is then defined with goals like minimizing the gate count, circuit depth, and improving fidelity, while tools like machine learning (ML), algorithms, and variational principles are employed to address the problem. The process involves iterative refinement to decode the optimized circuit structure, resulting in a more efficient quantum circuit ready for implementation. Figure 11. emphasizes systematic problem-solving and tool-assisted optimization to achieve high-performance circuits.

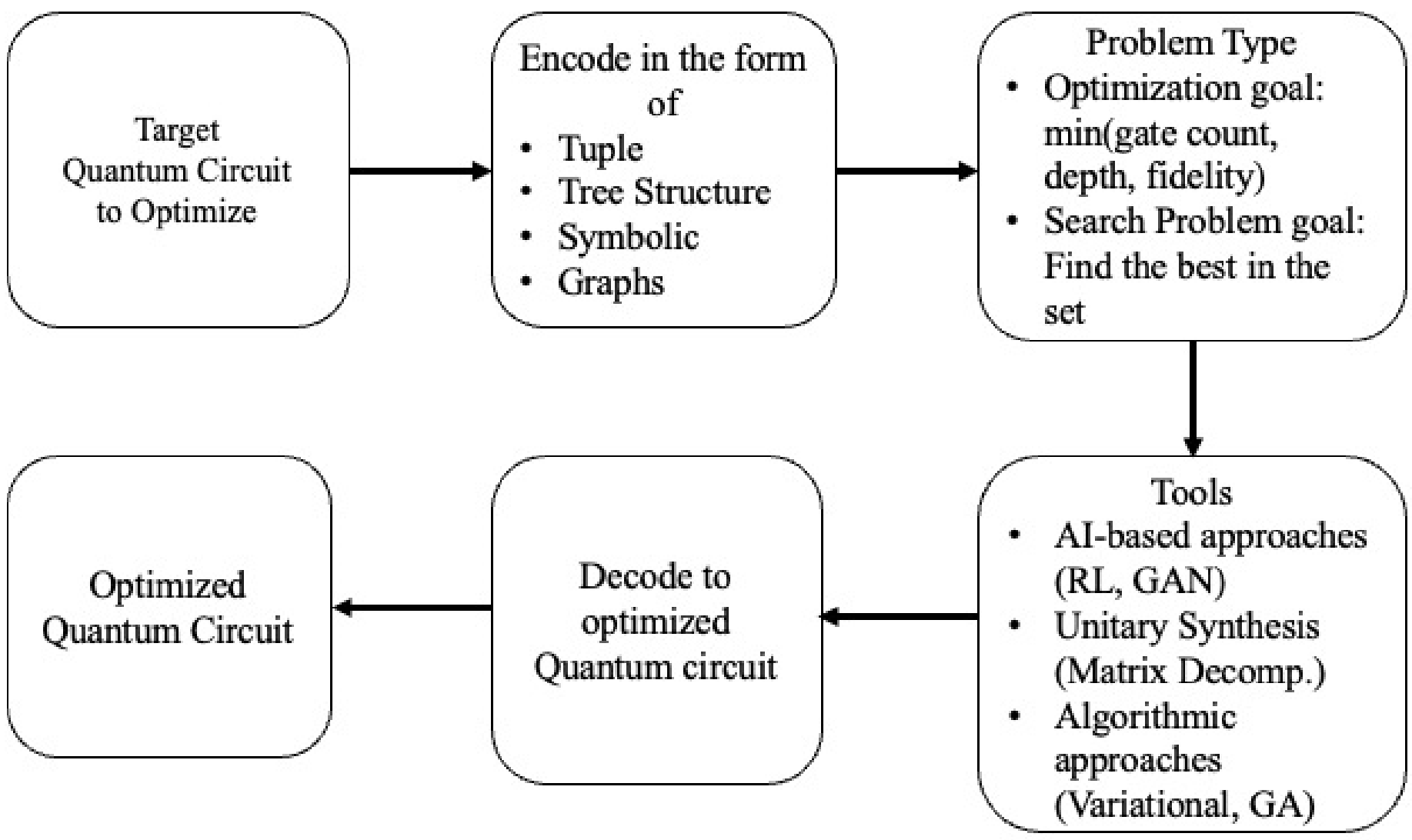

**Figure 11.** Summary of quantum circuity optimization methods.

Quantum circuit optimization has made significant advancements, yet several challenges remain, particularly in the Noisy Intermediate-Scale Quantum (NISQ) era. Noise and decoherence in NISQ devices, stemming from qubit susceptibility to gate errors and environmental factors, continue to degrade computation fidelity as circuit depth and gate count increase. Addressing this requires noise-aware optimization frameworks that integrate error-resilient gates and dynamically adapt to hardware error characteristics. Scalability is another critical issue, as many methods falter when applied to large circuits with thousands of gates or qubits, necessitating hybrid classical-quantum approaches, hierarchical frameworks, and adaptive strategies to balance accuracy and efficiency. The long-term goal of achieving fault-tolerant quantum computing depends on breakthroughs in error correction and optimization techniques to enable large-scale, high-fidelity computations. Additionally, universal optimization frameworks that combine hybrid, heuristic, and AI-based methods are vital for standardizing processes across diverse hardware platforms and algorithms, ensuring compatibility and scalability. By addressing these challenges, researchers can advance scalable, reliable, and efficient quantum systems, paving the way for practical quantum computing applications.

This paper provides a comprehensive literature review on the topic of quantum circuit optimization, highlighting the strategies and guidelines for enhancing circuit efficiency, as well as the implementation techniques for scaling up quantum circuits. The paper also touches on the distinction between logical quantum circuits and physical quantum circuits. To achieve efficiency goals in quantum computing, the current quantum era needs to use a combination of both technology-independent and technology-dependent optimization techniques. However, given the current limitations of scalable quantum computer technology, machine-independent optimization techniques are receiving more attention.

**Funding:** This research received no external funding.

**Data Availability Statement:** No new data were created or analyzed in this study.

**Conflicts of Interest:** The authors declare no conflicts of interest.